\documentclass[aps,pre,psfig,twocolumn,showpacs,superscriptaddress,reprint,floatfix]{revtex4-2}
\usepackage{amsmath}
\usepackage{mathtools}
\usepackage{amssymb}
\usepackage{color}
\usepackage{graphics}
\usepackage{epsfig}
\usepackage[normalem]{ulem} %to strike the words
\usepackage{cancel}
\usepackage[mathlines]{lineno}
\graphicspath{{Figures/}}
\usepackage{svg}
\usepackage{xtab,afterpage,longtable}
\usepackage{booktabs}   
\usepackage{ltablex}
\usepackage{enumitem}
\relax
\usepackage{float}
\usepackage{orcidlink}
\usepackage{hyperref}
\hypersetup{colorlinks=true,linkcolor=blue,urlcolor=blue,citecolor=blue}

\begin{document}
%%%%%%%%%%%%%%%%%%%%%%%%%%%%%%%%%%%%%%%%%%%%%%%%%%%%%%%%%%%%%%%%%%%%%%%%%%%%%%%%
%%%%%%%%%%%%%%%%%%%%%%%%%%%%%%%%%%%%%%%%%%%%%%%%%%%%%%%%%%%%%%%%%%%%%%%%%%%%%%%%
\title{Modeling of Experimentally Observed Two-Dimensional Precursor Solitons in a Dusty Plasma by the forced Kadomtsev–Petviashvili Equation}
\author{Ajaz Mir\orcidlink{0000-0001-5540-4967}}
\email{ajazmir.physics@gmail.com}
\affiliation{Institute for Plasma Research, Gandhinagar 382428, Gujarat, India} 
%\affiliation{Homi Bhabha National Institute, Anushaktinagar, Mumbai 400094, India}
\author{Pintu Bandyopadhyay\orcidlink{0000-0002-1857-8711}}
\email{pintu@ipr.res.in}
\affiliation{Institute for Plasma Research, Gandhinagar 382428, Gujarat, India} 
\affiliation{Homi Bhabha National Institute, Anushaktinagar, Mumbai 400094, India}
\author{Madhurima Choudhury\orcidlink{0000-0003-4395-4931}}
\affiliation{Brown University, 184 Hope Street, Providence, Rhode Island 02912, USA}
\author{Krishan Kumar\orcidlink{0000-0001-8287-7495}}
\affiliation{Department of Physics and Astronomy and the Center for KINETIC Plasma Physics, West Virginia University, Morgantown, West Virginia 26506, USA}
\author{Abhijit Sen\orcidlink{0000-0001-9878-4330}}
\affiliation{Institute for Plasma Research, Gandhinagar 382428, Gujarat, India} 
\affiliation{Homi Bhabha National Institute, Anushaktinagar, Mumbai 400094, India}
\date{\today}
%%%%%%%%%%%%%%%%%%%%%%%%%%%%%%%%%%%%%%%%%%%%%%%%%%%%%%%%%%%%%%%%%%%%%%%%%%%%%%%%
%%%%%%%%%%%%%%%%%%%%%%%%%%%%%%%%%%%%%%%%%%%%%%%%%%%%%%%%%%%%%%%%%%%%%%%%%%%%%%%%
\begin{abstract}
We compare model solutions of a forced Kadomtsev–Petviashvili (fKP) equation with experimental observations of dust acoustic precursor solitons excited by a supersonically moving charged cylindrical object in a dusty plasma medium. The fKP equation is derived from a three-fluid-Poisson model of the dusty plasma using the reductive perturbation technique and numerically solved for parameters close to the experimental investigations of cylindrical precursor solitons. The fKP model solutions show excellent agreement with the experimental results in reproducing the prominent geometric features of the two-dimensional solitons and closely matching the quantitative values of their velocities, amplitudes, and temporal evolutions.  Our findings suggest that the fKP equation can serve as a very realistic model to investigate the dynamics of precursor solitons and can be usefully employed in practical applications such as space debris detection and tracking techniques that are based on observing/predicting nonlinear plasma excitations induced by the debris in the ionosphere.  
\end{abstract}
%%%%%%%%%%%%%%%%%%%%%%%%%%%%%%%%%%%%%%%%%%%%%%%%%%%%%%%%%%%%%%%%%%%%%%%%%%%%%%%%
%%%%%%%%%%%%%%%%%%%%%%%%%%%%%%%%%%%%%%%%%%%%%%%%%%%%%%%%%%%%%%%%%%%%%%%%%%%%%%%%
\maketitle
%%%%%%%%%%%%%%%%%%%%%%%%%%%%%%%%%%%%%%%%%%%%%%%%%%%%%%%%%%%%%%%%%%%%%%%%%%%%%%%%
%%%%%%%%%%%%%%%%%%%%%%%%%%%%%%%%%%%%%%%%%%%%%%%%%%%%%%%%%%%%%%%%%%%%%%%%%%%%%%%%
\section{Introduction}
\label{intro}
%%%%%%%%%%%%%%%%%%%%%%%%%%%%%%%%%%%%%%%%%%%%%%%%%%%%%%%%%%%%%%%%%%%%%%%%%%%%%%%%
%%%%%%%%%%%%%%%%%%%%%%%%%%%%%%%%%%%%%%%%%%%%%%%%%%%%%%%%%%%%%%%%%%%%%%%%%%%%%%%%
\paragraph*{} 
The topic of precursor and pinned solitons excited by a fast-moving charged object in a plasma has received much attention lately after it was first suggested in~\cite{Sen_ASR_2015} that such solitons could potentially prove useful in the detection and tracking of small-sized space debris objects orbiting the Earth in the Low Earth Orbital (LEO) and Geosynchronous Earth Orbital (GEO) regions~\cite{Truitt_JSR_2020A, Truitt_JSR_2020B, Truitt_JSR_2021, Acharya_PRE_2021, Sen_POP_2023, Vikram_PRE_2023, Bernhardt_POP_2023}. These objects, which are difficult to detect optically and possess enormous kinetic energy, can cause serious damage to active space assets through their collisional impacts. Their detection and tracking are therefore of paramount importance and continue to be a major area of research in the Space Situational Awareness program. One characteristic feature of space debris objects is that they get highly charged through the collection of electrons and ions from the surrounding plasma and a variety of other processes like the production of photo-electrons due to ultraviolet and x-rays from the sun and the emission of secondary electrons and ions from the surface~\cite{Lai_IEEE_2012, Sen_ASR_2015, Sanat_AA_2024}. As charged objects, they can then interact electromagnetically with the plasma to create a variety of linear and nonlinear waves. Under certain conditions, dependent on the speed and amount of charge on the object, they can give rise to coherent nonlinear structures like precursor solitons. 
%which can travel ahead of them at a faster speed. 
\textcolor{black}{ Precursor solitons are nonlinear localized wave structures that travel faster than the source exciting them and propagate for a long distance. They satisfy the mathematical condition of the constancy of the quantity $AW^2$, where $A$ is the amplitude and $W$ is the width of the structure. Precursor solitons have been extensively studied in the context of the one-dimensional forced Korteweg-de Vries (fKdV) equation to model such excitations in hydrodynamics as well as in plasmas~\cite{Sen_ASR_2015, Surabhi_PRE_2016, Truitt_JSR_2020A, Wu_JFM_1987, Lee_JFM_1989}. The precursor solitons of the forced Kadomtsev-Petviashvili (fKP) equation, discussed in the present work, are the two-dimensional generalization of the fKdV solitons that possess a transverse structure that gives them a cylindrical or spherical shell shape.}
The detection and tracking of such nonlinear plasma signatures, which can have a larger footprint than the size of the object, formed the basis for the technique suggested in~\cite{Sen_ASR_2015}. It was further shown that a simple one-dimensional (1D) model equation in the form of the fKdV equation was capable of capturing the basic features of the excitation and propagation of precursor solitons in the plasma and could be usefully employed to develop this scheme further. Subsequently, the fKdV equation was extensively investigated by Truitt~{\it et al.}~\cite{Truitt_JSR_2020A, Truitt_JSR_2020B} to explore the feasibility of the scheme for ionospheric conditions. To account for higher dimensional effects a two-dimensional (2D) generalization of the fKdV equation in the form of the fKP equation was also numerically investigated by Truitt~{\it et al.}~\cite{Truitt_JSR_2021} in different parameter regimes. 
\paragraph*{}
While the fKdV and fKP serve as excellent mathematical models for a qualitative understanding of the basic features of precursor soliton excitation and propagation characteristics, it is necessary to validate their applicability in practical situations by comparing their solutions with experimental results. This is the primary motivation of the present paper, where we test the validity of the fKP equation in quantitatively modeling the experimental results of cylindrical precursor solitons obtained in a controlled laboratory setting. These experiments were carried out in a dusty plasma medium. A dusty plasma is a mixture of electrons, positively charged ions, neutral particles, and heavily charged (primarily negative)~\cite{Barkan_PRL_1994} dust grains. They serve as a convenient test bed for studying low-frequency linear and nonlinear wave phenomena in plasmas due to the convenience of visually capturing the slow motion of the massive dust particles and the concomitant wave motion using video recordings~\cite{Surabhi_RSI_2015}. In fact, the first experimental demonstration of precursor solitons was obtained in a dusty plasma in which the dust fluid was made to flow supersonically over a charged wire~\cite{Surabhi_PRE_2016}. The electrostatic potential hill arising from the sheath formed around the wire served as a charged obstacle, and when the dust flow exceeded a certain value, precursor solitons were observed to be excited. The nature of the excited solitary structures was confirmed to be solitons by establishing that the quantity $AW^2$ remained a constant, where $A$ represents the amplitude and $W$ the width of the structure. It should be mentioned that the process of precursor excitation is Galilean invariant. In other words, it is independent of whether the charged source is made to move through a fluid or the fluid is made to move over a stationary object. Experimentally, it is easier to make the fluid flow in a controlled manner and has become the preferred approach in experiments.  Subsequent experiments were able to successfully reproduce the early results and also extend the investigations to examine the effects of the size and shape of the 1D source and look at other nonlinear structures like pinned solitons~\cite{Garima_PRE_2021, Garima_POP_2019}.
\paragraph*{}
More recent experiments have explored the nature of the precursor solitons when the source is of a two-dimensional (cylindrical) shape or a three-dimensional (spherical) shape. The fKdV equation, being of a one-dimensional nature, is inadequate to describe such higher-dimensional excitations. We, therefore, consider the two-dimensional fKP equation and test its validity in both qualitatively and quantitatively interpreting the dust acoustic (DA) precursor soliton results obtained by Kumar~{\it et al.}~\cite{Krishan_POP_2024} for cylindrical solitons. For this, we first derive an appropriate fKP equation for a dusty plasma from a three-fluid-Poisson model by means of the reductive perturbation method. 
%The equation is then bench-marked by numerically solving it for different kinds of driving sources and examining the nature of the resultant solutions. 
The equation is then numerically solved using parameters taken from the experimental conditions, and the numerical results are compared to the experimental ones. It is found that the fKP results agree remarkably well with the experimental ones, both in their qualitative features and in their quantitative measures.
\textcolor{black}{ Our findings indicate that within the constraints of the fluid description, the fKP model provides a very realistic model for characterizing the phenomenon of excitation of precursor solitons in plasma and may be usefully employed in practical applications such as in the context of small debris detection and tracking.}
\paragraph*{}
The manuscript is organized as follows. 
The fluid model adopted to explain the electrostatic excitations in a dusty plasma medium is described in section~\ref{DP_MODEL}. This section also contains a derivation of the fKP equation using the reductive perturbation technique. Section~\ref{fKP_Exp} provides a detailed comparison of the fKP numerical results with those of experimental observations. Section~\ref{sum_con} contains a summary of our results, their implications for practical applications of the fKP model, and a discussion of the limitations of the model and possible future improvements. 
%%%%%%%%%%%%%%%%%%%%%%%%%%%%%%%%%%%%%%%%%%%%%%%%%%%%%%%%%%%%%%%%%%%%%%%%%%%%%%%%
%%%%%%%%%%%%%%%%%%%%%%%%%%%%%%%%%%%%%%%%%%%%%%%%%%%%%%%%%%%%%%%%%%%%%%%%%%%%%%%%
\section{Derivation of the forced Kadomtsev-Petviashvili evolution equation for a dusty plasma}
\label{DP_MODEL}
%%%%%%%%%%%%%%%%%%%%%%%%%%%%%%%%%%%%%%%%%%%%%%%%%%%%%%%%%%%%%%%%%%%%%%%%%%%%%%%%
%%%%%%%%%%%%%%%%%%%%%%%%%%%%%%%%%%%%%%%%%%%%%%%%%%%%%%%%%%%%%%%%%%%%%%%%%%%%%%%%
\textcolor{black}{
We consider the standard fluid model for a dusty plasma that assumes the ions and electrons to be inertial-less compared to the heavy dust particles and can be described by Maxwell-Boltzmann distributions ~\cite{Sanat_NJP_2012, Shukla_IOP_2001, Rao_PSS_1990}
%%%%%%%%%%%%%%%%%%%%%%%%%%%%%%%%%%%%%%%%%
\begin{equation}\label{BEI_eqn4}
    n_{e, i} = n_{e0, i0} \exp(\pm e\phi / k_B T_{e,i}) .
\end{equation}
%%%%%%%%%%%%%%%%%%%%%%%%%%%%%%%%%%%%%%%%%
The subscripts e and i refer to electron and ion species, `0' refers to the equilibrium quantity, and the $+$ sign in the exponential is for the electrons and the $-$ for the ions. The equilibrium densities satisfy the quasi-neutrality condition $n_{i0} = Z_d n_{0} + n_{e0}$, where $Z_d$ is the dust charge number, and $n_{0}$ is the equilibrium dust density. The equations describing the dynamics of the dust fluid consist of the continuity equation and the momentum equation. The continuity equation in normalized form is given by
%%%%%%%%%%%%%%%%%%%%%%%%%%%%%%%%%%%%%%%%%%%%%%%%%%%
\begin{equation}\label{NCE_eqn1}
\frac{\partial n}{\partial t} 
+
\frac{\partial (n u_x)}{\partial x} 
+
\frac{\partial (n u_y)}{\partial y}
= 0.
\end{equation}
%%%%%%%%%%%%%%%%%%%%%%%%%%%%%%%%%%%%%%%%%%%%%%%%%%%
\textcolor{black}{where $x$ and $y$ are the normalized spatial coordinates along the direction of propagation of the source and in the transverse direction, respectively.}
The x and y components of the dust momentum equation~\textcolor{black}{in normalized form} are given by
%%%%%%%%%%%%%%%%%%%%%%%%%%%%%%%%%%%%%%%%%%%%%%%%%%%
\begin{equation}\label{NMEX_eqn2}
 \frac{\partial u_x}{\partial t} 
+ u_x \frac{\partial u_x}{\partial x}
+ u_y \frac{\partial u_x}{\partial y}
= \frac{\partial\phi}{\partial x}  ,
\end{equation}
%%%%%%%%%%%%%%%%%%%%%%%%%%%%%%%%%%%%%%%%%%%%%%%%%%%
\begin{equation}\label{NMEY_eqn2}
 \frac{\partial u_y}{\partial t} 
+ u_x \frac{\partial u_y}{\partial x}
+ u_y \frac{\partial u_y}{\partial y}
= \frac{\partial\phi}{\partial y}   .
\end{equation}
%%%%%%%%%%%%%%%%%%%%%%%%%%%%%%%%%%%%%%%%%%%%%%%%%%%
The electrostatic fluctuations of the above governing equations are closed by Poisson’s equation~\textcolor{black}{in normalized form} given by
%%%%%%%%%%%%%%%%%%%%%%%%%%%%%%%%%%%%%%%%%%%%%%%%%%%
\textcolor{black}{
\begin{eqnarray}\label{NPE_eqn3}
 \frac{\partial^2 \phi}{\partial x^2} 
+ \frac{\partial^2 \phi}{\partial y^2}
\nonumber
    = n + \mu_e e^{\sigma_i \phi} - \mu_i e^{-\phi} 
    \\ \nonumber
    +\ S(x - v_{dx} t, y - v_{dy} t) 
    \\
    \approx n  - 1 + c_1 \phi + c_2 \phi^2 + ... 
    \\ \nonumber
    +\ S(x - v_{dx} t, y - v_{dy} t) ,
\end{eqnarray} }
%%%%%%%%%%%%%%%%%%%%%%%%%%%%%%%%%%%%%%%%%%%%%%%%%%%
}
where $S$ represents the charge density of the external source (the charged debris object) that is moving with respect to the plasma with a velocity $(v_{dx},v_{dy})$. Other quantities are defined as $\mu_e =\ n_{e0}/ (Z_d n_{d0})$; $\mu_i =\ n_{i0}/ (Z_d n_{d0})$; and $\sigma_i =\ T_i/T_e$. Also, $c_1 = \mu_e \sigma_i + \mu_i$ and $c_2 = (\mu_e \sigma_i^2 - \mu_i)/2$.
The dependent variables $n(x,y)$, $u_x(x,y), u_y(x,y)$, $\phi(x,y)$ and the independent variables $x,y,t$ are normalized as follows. The space, time, velocity, density, and electrostatic potential are normalized as  $x = x/\lambda_D$, $t = t \omega_{pd}$, $u_x = u_{x}/\lambda_D \omega_{pd}$, $u_y = u_{y}/\lambda_D \omega_{pd}$, $n = n_d/n_{d0};\ n_i = n_i/n_{i0};\ n_e = n_e/n_{e0}$ and $\phi = e\phi/ k_B T_i $, respectively. The dust plasma frequency is defined as $\omega_{pd} = \sqrt{n_{d0} (Z_d e)^2/ \epsilon_0 m_d}$. The  dust Debye length  $\lambda_{D} =\ \sqrt{\epsilon_0 k_B T_i/ n_{d0} Z_d e^2}$ makes the coefficient of the $\partial \phi / \partial x$ term in \eqref{NMEX_eqn2} and $\partial \phi / \partial y$ \eqref{NMEY_eqn2} unity.
The normalized velocity of the source $v_{dx} = v_{dx}/\lambda_D \omega_{pd}$ and $v_{dy} = v_{dx}/\lambda_D \omega_{pd}$. 
\paragraph*{}
To derive the nonlinear evolution equation, we employ the standard reductive perturbation technique~\cite{Sen_ASR_2015, Acharya_PRE_2021} and expand the dynamical quantities in a power series based on the small dimensionless expansion parameter $\epsilon$, which quantifies the system's nonlinearity strength.
%%%%%%%%%%%%%%%%%%%%%%%%%%%%%%%%%%%%%%%%%%%%%%%%%%%
\begin{equation}\label{density_1}
    n = 1 + \epsilon n_1 + \epsilon^2 n_2 + \epsilon^3 n_3 + O(\epsilon^4)
\end{equation}
\begin{equation}\label{vel_x1}
    u_x =  \epsilon u_{x_1} + \epsilon^2 u_{x_2}  + \epsilon^3 u_{x_3}  + O(\epsilon^4)
\end{equation}
\begin{equation}\label{vel_y1}
    u_y =  \epsilon^{3/2} u_{y_1} + \epsilon^{5/2} u_{y_2}  + \epsilon^{7/2} u_{y_3}  + O(\epsilon^{9/2})
\end{equation}
\begin{equation}\label{potential_1}
    \phi =  \epsilon \phi_1 + \epsilon^2 \phi_2 + \epsilon^3 \phi_3 + O(\epsilon^4) .
\end{equation}
%%%%%%%%%%%%%%%%%%%%%%%%%%%%%%%%%%%%%%%%%%%%%%%%%%%
We consider a weak, localized, two-dimensional space-time dependent charge density source function vanishing at infinity that is given by
%%%%%%%%%%%%%%%%%%%%%%%%%%%%%%%%%%%%%%%%%%%%%%%%%%%
\begin{equation}\label{source_1}
    S(x,y,t) = \epsilon^2 S_2(x,y,t).
\end{equation}
\noindent
%%%%%%%%%%%%%%%%%%%%%%%%%%%%%%%%%%%%%%%%%%%%%%%%%%%
\textcolor{black}{The forcing term in the fKP equation physically represents the charged object in the experiment. It is a moving source in the frame of the flowing dust fluid and is responsible for the excitation of various perturbations in the plasma in the form of wakes and solitons. 
In mathematically deriving the fKP equation within the framework of the reductive perturbation method, the forcing term is introduced at the appropriate order of $\epsilon$ to ensure that it balances with the nonlinear and dispersive terms in the equation.
Such an ordering is in keeping with the weakly nonlinear and weakly dispersive approximation for the system. The external source (forcing) is taken to be small enough to match the scales of nonlinearity and dispersion, and strong enough to generate coherent excitations (\textit{e.g.,} solitary or wake structures).}
\paragraph*{}
\textcolor{black}{Mathematically, the power of the expansion parameter $\epsilon$ defines the order of approximation adopted to describe the dynamics of the system. In deriving the KP equation, it is chosen such that there is a balance of the nonlinearity and dispersion in the system that can lead to the formation of nonlinear structures like solitons. 
In physical terms, the parameter $\epsilon$ is a measure of the smallness of the perturbed amplitude of a physical variable to its corresponding equilibrium value (\textit{i.e.,} $n_1/n_0 \ll 1$,  where $n_1$ is the perturbed density and $n_0$ is the equilibrium density). It also quantifies the smallness of the dispersion contribution (\textit{i.e.,} $k\lambda_{D}\ll 1$, where $k$ is the wave number).
}
%\paragraph*{}
%held fixed when analyzing the temporal evolution of the system at specific spatial locations, particularly within the framework of the multiple scales method, where variations in space and time are separated as described above.
%This distinction helps to isolate the time-dependent behavior of the charge density and potential fields at given spatial points.
\paragraph*{}
We next introduce a set of stretched coordinates defined as follows:
%%%%%%%%%%%%%%%%%%%%%%%%%%%%%%%%%%%%%%%%%%%%%%%%%%%
\begin{equation}\label{stretch_1}
    \xi = \epsilon^{1/2} (x - v_{ph} t); 
    \hspace{0.5cm} \eta = \epsilon y; 
    \hspace{0.5cm} \tau = \epsilon^{3/2} t
\end{equation}
%%%%%%%%%%%%%%%%%%%%%%%%%%%%%%%%%%%%%%%%%%%%%%%%%%%
where $v_{ph}$ is the phase velocity of the wave. The differential operators used in the above governing equations in terms of these stretched variables are then given by,
%%%%%%%%%%%%%%%%%%%%%%%%%%%%%%%%%%%%%%%%%%%%%%%%%%%%%%%%
\begin{eqnarray}\label{stretch_2}
  \frac{\partial}{\partial x} = \epsilon^{1/2} \frac{\partial }{\partial \xi};
  \hspace{1cm}
  \frac{\partial}{\partial y} = \epsilon \frac{\partial }{\partial \eta};
  \\ \nonumber
   \frac{\partial}{\partial t}  = - v_{ph} \epsilon^{1/2}\frac{\partial }{\partial \xi} + \epsilon^{3/2} \frac{\partial }{\partial \tau} .
\end{eqnarray}
%%%%%%%%%%%%%%%%%%%%%%%%%%%%%%%%%%%%%%%%%%%%%%%%%%%%%%%%
The two time scales distinguish between the dynamics of the fast (or lowest limit of $\epsilon$) and the slow (or the next higher limit of $\epsilon$) time scales. Physically, the perturbation follows the linear wave equation at a fast dynamical time scale and propagates with the dust acoustic speed. In contrast, at a slow dynamical time scale, the waveform is influenced collectively by weak nonlinear steepening and dispersive broadening. Due to the balance between these two effects, the perturbation forms a solitary wave structure. 
\paragraph*{}
Substituting the expansions for $n$, $u_x$, $u_y$, and $\phi$ in Eqns.~\eqref{NCE_eqn1}-\eqref{NPE_eqn3} and expressing the space-time differential operators in terms of the stretched variables, we get,
%%%%%%%%%%%%%%%%%%%%%%%%%%%%%%%%%%%%%%%%%%%%%%%%%%%%%%%%
\begin{equation}\label{exp_CE_eqn1}
        \begin{rcases}
            \left( -v_{ph}\epsilon^{1/2} \frac{\partial}{\partial \xi} + \epsilon^{3/2} \frac{\partial}{\partial \tau}\right) (1+\epsilon n_{1}+\epsilon^2 n_{2}+...) 
            \\
            +
            \epsilon^{1/2} \frac{\partial }{\partial \xi} (1+\epsilon n_{1}+\epsilon^2 n_{2}+...) (\epsilon u_{x_1}+\epsilon^2 u_{x_2} +...)
            \\
            +
            \epsilon \frac{\partial }{\partial \eta} (1+\epsilon n_{1}+\epsilon^2 n_{2}+...) (\epsilon^{3/2} u_{y_1}+\epsilon^{5/2}u_{y_2}+...)
            \\
            =0
        \end{rcases}
\end{equation}
%%%%%%%%%%%%%%%%%%%%%%%%%%%%%%%%%%%%%%%%%%%%%%%%%%%
\begin{equation}\label{exp_xME_eqn2}
        \begin{rcases}
           \left( -v_{ph} \epsilon^{1/2} \frac{\partial}{\partial \xi} + \epsilon^{3/2} \frac{\partial}{\partial \tau}\right)  (\epsilon u_{x_1}+\epsilon^2 u_{x_2}+...)
            \\
           +
           (\epsilon u_{x_1}+\epsilon^2 u_{x_2} +...) \epsilon^{1/2} \frac{\partial }{\partial \xi}  (\epsilon u_{x_1}+\epsilon^2 u_{x_2} +...) 
           \\
           +
           (\epsilon^{3/2} u_{y_1}+\epsilon^{5/2}u_{y_2}+...)\epsilon \frac{\partial }{\partial \eta}(\epsilon u_{x_1}+\epsilon^2 u_{x_2} +...)
           \\
           -
           \epsilon^{1/2} \frac{\partial }{\partial \xi}(\epsilon \phi_1 + \epsilon^2 \phi_2  + ...)
           = 0
        \end{rcases}
\end{equation}
%%%%%%%%%%%%%%%%%%%%%%%%%%%%%%%%%%%%%%%%%%%%%%%%%%%
\begin{equation}\label{exp_yME_eqn2}
        \begin{rcases}
            \left( -v_{ph} \epsilon^{1/2} \frac{\partial}{\partial \xi} + \epsilon^{3/2} \frac{\partial}{\partial \tau}\right) (\epsilon^{3/2} u_{y_1}+\epsilon^{5/2}u_{y_2}+...)
            \\
            +
            (\epsilon u_{x_1}+\epsilon^2 u_{x_2} +...)\epsilon^{1/2} \frac{\partial }{\partial \xi}(\epsilon^{3/2} u_{y_1}+\epsilon^{5/2} u_{y_2} +...)
            \\
            +
            (\epsilon^{3/2} u_{y_1}+\epsilon^{5/2} u_{y_2} +...)\epsilon \frac{\partial }{\partial \eta}
            (\epsilon^{3/2} u_{y_1}+\epsilon^{5/2} u_{y_2} +...)
            \\
            -
            \epsilon \frac{\partial }{\partial \eta}(\epsilon \phi_1 + \epsilon^2 \phi_2  + ...)
            = 0
        \end{rcases}
\end{equation}
%%%%%%%%%%%%%%%%%%%%%%%%%%%%%%%%%%%%%%%%%%%%%%%%%%%
\begin{equation}\label{exp_PE_eqn3}
        \begin{rcases}
            \epsilon\frac{\partial^2}{\partial \xi^2}(\epsilon\phi_1 +\epsilon^2 \phi_2 + ...)
            +
            \epsilon^2 \frac{\partial^2}{\partial \eta^2}(\epsilon\phi_1 +\epsilon^2 \phi_2 + ...)
            \\
            -
            (1 +\ \epsilon n_1 +\ \epsilon^2 n_2 +\ ...)
            \\
            +
            1 
            -
           c_1 (\epsilon \phi_1 + \epsilon^2 \phi_2 + ...) 
           -
           c_2 (\epsilon \phi_1 + \epsilon^2 \phi_2 + ...)^2 
           \\
        = 
        \epsilon^2 S_2(\xi + (v_{ph} - v_d)\tau, \eta + (v_{ph} - v_d)\tau)
        \end{rcases}
\end{equation}
%%%%%%%%%%%%%%%%%%%%%%%%%%%%%%%%%%%%%%%%%%%%%%%%%%%%%%%%
Equating the same order terms in  $\epsilon$ from \eqref{exp_CE_eqn1}, we get
%%%%%%%%%%%%%%%%%%%%%%%%%%%%%%%%%%%%%%%%%%%%%%%%%%%
\begin{equation}\label{CE_3by2_eq1}
    \epsilon^{3/2}: 
    - v_{ph} \frac{\partial n_1}{\partial \xi} 
    + \frac{\partial u_{x_1}}{\partial \xi} 
    = 0 
\end{equation}
%%%%%%%%%%%%%%%%%%%%%%%%%%%%%%%%%%%%%%%%%%%%%%%%%%%
\begin{equation}\label{CE_5by2_eq1}
    \epsilon^{5/2}: 
    - v_{ph} \frac{\partial n_2}{\partial \xi} 
    + \frac{\partial n_1}{\partial \tau}
    + \frac{\partial n_1 u_{x_1}}{\partial \xi} 
    +\frac{\partial u_{x_2}}{\partial \xi} 
    + \frac{\partial u_{y_1}}{\partial \eta} 
    = 0 
\end{equation}
%%%%%%%%%%%%%%%%%%%%%%%%%%%%%%%%%%%%%%%%%%%%%%%%%%%
Equating the same order terms in  $\epsilon$ from \eqref{exp_xME_eqn2}, we get
%%%%%%%%%%%%%%%%%%%%%%%%%%%%%%%%%%%%%%%%%%%%%%%%%%%
\begin{equation}\label{xME_3by2_eq2}
    \epsilon^{3/2}: 
    - v_{ph} \frac{\partial u_{x_1}}{\partial \xi} 
    - \frac{\partial \phi_{1}}{\partial \xi} 
    = 0 
\end{equation}
%%%%%%%%%%%%%%%%%%%%%%%%%%%%%%%%%%%%%%%%%%%%%%%%%%%
\begin{equation}\label{xME_5by2_eq2}
    \epsilon^{5/2}: 
    - v_{ph} \frac{\partial u_{x_2}}{\partial \xi} 
    + \frac{\partial u_{x_1}}{\partial \tau}
    + u_{x_1}\frac{\partial u_{x_1}}{\partial \xi} 
    - \frac{\partial \phi_{2}}{\partial \xi} 
    = 0 
\end{equation}
%%%%%%%%%%%%%%%%%%%%%%%%%%%%%%%%%%%%%%%%%%%%%%%%%%%
Equating the same order terms in  $\epsilon$ from \eqref{exp_yME_eqn2}, we get
%%%%%%%%%%%%%%%%%%%%%%%%%%%%%%%%%%%%%%%%%%
\begin{equation}\label{yME_2_eq2}
    \epsilon^{2}: 
    - v_{ph} \frac{\partial u_{y_1}}{\partial \xi}                               
    - \frac{\partial \phi_{1}}{\partial \eta} 
    = 0 
\end{equation}
%%%%%%%%%%%%%%%%%%%%%%%%%%%%%%%%%%%%%%%%%%%%%%%%%%%
%%%%%%%%%%%%%%%%%%%%%%%%%%%%%%%%%%%%%%%%%%%%%%%%%%%
Equating the same order terms in  $\epsilon$ from \eqref{exp_PE_eqn3}, we get
%%%%%%%%%%%%%%%%%%%%%%%%%%%%%%%%%%%%%%%%%%
\begin{equation}\label{PE_1_eq3} 
  \epsilon^{1}: 
    - c_1 \phi_1 - n_1 = 0 \Rightarrow n_1 = - c_1 \phi_1 %;\ c_1 = 1
\end{equation}
%%%%%%%%%%%%%%%%%%%%%%%%%%%%%%%%%%%%%%%%%%%%%%%%%%%
\begin{equation}\label{PE_2_eq3}
    \epsilon^{2}: 
    \frac{\partial^2 \phi_1}{\partial \xi^2} - n_2 - c_1 \phi_2 - c_2 \phi_1^2 
    = S_2 
\end{equation}
%%%%%%%%%%%%%%%%%%%%%%%%%%%%%%%%%%%%%%%%%%%%%%%%%%%
%%%%%%%%%%%%%%%%%%%%%%%%%%%%%%%%%%%%%%%%%%%%%%%%%%%
From \eqref{CE_3by2_eq1},  \eqref{xME_3by2_eq2},  and \eqref{PE_1_eq3}, we get
%%%%%%%%%%%%%%%%%%%%%%%%%%%%%%%%%%%%%%%%%%%%%%%%%%%
\begin{equation}
    n_1 = u_{x_1}/v_{ph};\ u_{x_1} = - \phi_1 / v_{ph} ;\ v_{ph} = \pm 1/\sqrt{c_1}
\end{equation}
%%%%%%%%%%%%%%%%%%%%%%%%%%%%%%%%%%%%%%%%%%%%%%%%%%%
From \eqref{yME_2_eq2}, we get
%%%%%%%%%%%%%%%%%%%%%%%%%%%%%%%%%%%%%%%%%%%%%%%%%%%
\begin{equation}\label{vy1_phi1}
    \frac{\partial u_{y_1}}{\partial \xi} = - \frac{1}{v_{ph}} \frac{\partial \phi_1}{\partial \eta} %;\ v_{ph} = 1   
\end{equation}
%%%%%%%%%%%%%%%%%%%%%%%%%%%%%%%%%%%%%%%%%%%%%%%%%%%
Multiplying \eqref{CE_5by2_eq1} by $v_{ph}$, and substituting for $n_1$, and $v_{x_1}$ in terms of $\phi_1$ in
\eqref{CE_5by2_eq1} and \eqref{xME_5by2_eq2} and adding the two equations, we get
%%%%%%%%%%%%%%%%%%%%%%%%%%%%%%%%%%%%%%%%%%%%%%%%%%%
\begin{eqnarray}\label{add_5b2_CE_5by2_xME}
    -\left( c_1 v_{ph} + \frac{1}{v_{ph}} \right)   \frac{\partial \phi_1}{\partial \tau} 
 + \left(2c_1 + \frac{1}{v_{ph}^2} \right)  \phi_1 \frac{\partial \phi_1}{\partial \xi} 
 \\ \nonumber
    + \frac{\partial v_{y_1}}{\partial \eta} 
    = v_{ph}^2\frac{\partial n_2}{\partial \xi} + \frac{\partial \phi_2}{\partial \xi} 
\end{eqnarray}
%%%%%%%%%%%%%%%%%%%%%%%%%%%%%%%%%%%%%%%%%%%%%%%%%%%
Taking the derivative of \eqref{PE_2_eq3} w.r.t. $\xi$, we get
%%%%%%%%%%%%%%%%%%%%%%%%%%%%%%%%%%%%%%%%%%%%%%%%%%%
\begin{equation}\label{der_PE_3}
   v_{ph}^2 \frac{\partial n_2}{\partial \xi} + \frac{\partial \phi_2}{\partial \xi} 
    = 
   v_{ph}^2 \frac{\partial^3 \phi_1}{\partial \xi^3} 
    - 2c_2 v_{ph}^2 \phi_1 \frac{\partial \phi_1}{\partial \xi}
    - v_{ph}^2 \frac{\partial S_2}{\partial \xi}
\end{equation}
%%%%%%%%%%%%%%%%%%%%%%%%%%%%%%%%%%%%%%%%%%%%%%%%%%%
Substituting \eqref{der_PE_3} in \eqref{add_5b2_CE_5by2_xME}, we get
%%%%%%%%%%%%%%%%%%%%%%%%%%%%%%%%%%%%%%%%%%%%%%%%%%%
\begin{eqnarray}\label{add_5b2_CE_5by2_xME_1}
\left( c_1 v_{ph} + \frac{1}{v_{ph}} \right)  \frac{\partial \phi_1}{\partial \tau} 
\\ \nonumber
-\left(2c_1 + \frac{1}{v_{ph}^2} + 2c_2 v_{ph}^2 \right)  \phi_1 \frac{\partial \phi_1}{\partial \xi} 
\\ \nonumber
    + v_{ph}^2 \frac{\partial^3 \phi_1}{\partial \xi^3} 
    - v_{ph} \frac{\partial v_{y_1}}{\partial \eta} 
    = v_{ph}^2 \frac{\partial S_2}{\partial \xi}
\end{eqnarray}
%%%%%%%%%%%%%%%%%%%%%%%%%%%%%%%%%%%%%%%%%%%%%%%%%%%
Taking the derivative of \eqref{add_5b2_CE_5by2_xME_1} w.r.t. $\xi$, we get
%%%%%%%%%%%%%%%%%%%%%%%%%%%%%%%%%%%%%%%%%%%%%%%%%%%
% \begin{eqnarray}\label{fKP_0}
% \nonumber
% &    \frac{\partial}{\partial \xi} 
%       \left[
%      \left( \frac{2}{v_{ph}} \right) \frac{\partial \phi_1}{\partial \tau} 
%     - \left( \frac{3 + 2c_2 v_{ph}^4}{v_{ph}^2} \right) \phi_1 \frac{\partial \phi_1}{\partial \xi} 
%     + v_{ph}^2  \frac{\partial^3 \phi_1}{\partial \xi^3} 
%     \right]
%     \\ 
% &  - v_{ph} \frac{\partial}{\partial \xi} \left( \frac{\partial u_{y_1}}{\partial \eta} \right)
%    = v_{ph}^2 \frac{\partial^2 S_2}{\partial \xi^2} 
% \end{eqnarray}
%%%%%%%%%%%%%%%%%%%%%%%%%%%%%%%%%%%%%%%%%%%%%%%%%%%
\begin{flalign}\label{fKP_0}
\nonumber
&    \frac{\partial}{\partial \xi} 
      \left[
     \left( \frac{2}{v_{ph}} \right) \frac{\partial \phi_1}{\partial \tau} 
    - \left( \frac{3 + 2c_2 v_{ph}^4}{v_{ph}^2} \right) \phi_1 \frac{\partial \phi_1}{\partial \xi} 
    + v_{ph}^2  \frac{\partial^3 \phi_1}{\partial \xi^3} 
    \right]
    \\ 
&  - v_{ph} \frac{\partial}{\partial \xi} \left( \frac{\partial u_{y_1}}{\partial \eta} \right)
   = v_{ph}^2 \frac{\partial^2 S_2}{\partial \xi^2} 
\end{flalign}
%%%%%%%%%%%%%%%%%%%%%%%%%%%%%%%%%%%%%%%%%%%%%%%%%%%
After using~\eqref{vy1_phi1}  in~\eqref{fKP_0}, one gets the following two-dimensional nonlinear evolution equation
%%%%%%%%%%%%%%%%%%%%%%%%%%%%%%%%%%%%%%%%%%%%%%%%%%%
\begin{eqnarray}\label{fKP_equation2}
     \frac{\partial}{\partial \xi} 
   \left[
       \frac{\partial \phi_1}{\partial \tau} 
    -\ A_1 \phi_1 \frac{\partial \phi_1}{\partial \xi} 
    +\ A_2 \frac{\partial^3 \phi_1}{\partial \xi^3} 
    \right]
    \\ \nonumber
    +\ A_3 \frac{\partial^2 \phi_{1}}{\partial \eta^2} 
   =\ A_2 \frac{\partial^2 S_2}{\partial \xi^2}  
\end{eqnarray}
%%%%%%%%%%%%%%%%%%%%%%%%%%%%%%%%%%%%%%%%%%%%%%%%%%%
where $A_1 = (3 + 2c_2 v_{ph}^4 )/2v_{ph}$; $A_2 = v_{ph}^3/2$; $A_3 = v_{ph}/2$.
\paragraph*{}
In the absence of the source term, the L.H.S. of Eq.~\eqref{fKP_equation2} is the well-known Kadomtsev–Petviashvili (KP) equation,  a two-dimensional prototype nonlinear evolution equation which governs the dynamics of weakly nonlinear waves in the long-wavelength regime in various dispersive systems. The KP equation was first introduced by Kadomtsev and Petviashvili~\citep{Kadomtsev_SPD_1970} to study the stability of a Korteweg-de Vries soliton under transverse perturbations in a two-dimensional domain. The KP equation has been extensively employed in the past in a variety of situations \textit{e.g.,} to describe the nonlinear evolution of 2D solitons in hydrodynamics~\citep{Cho_PRE_2018}, in plasmas~\citep{Frycz_PRL_1989, Infeld_PRL_1994, Senatorski_PRE_1996} and diverse other nonlinear media~\cite{Baronio_PRL_2016, Cosme_PRB_2023}. The KP equation can be viewed as a 2D extension of the KdV equation, which takes into account the effect of weak transverse perturbations on the KdV-type solitary waves. Versions of the KP equation have been used in the past to model the propagation of nonlinear DA solitary waves in dusty plasmas in the presence of variable dust charge and two-temperature ions, as well as super-thermal electrons and ions ~\cite{Hamid_CSF_2009, Davoud_POP_2012, Saini_ASR_2015}. Therefore, one can expect the fKP equation \textit{i.e.,} Eq.~\eqref{fKP_equation2} to be an appropriate model to describe the excitation and propagation of precursor solitons in a driven dusty plasma medium.
%%%%%%%%%%%%%%%%%%%%%%%%%%%%%%%%%%%%
\begin{figure*} [ht!]
\includegraphics[width = \textwidth]{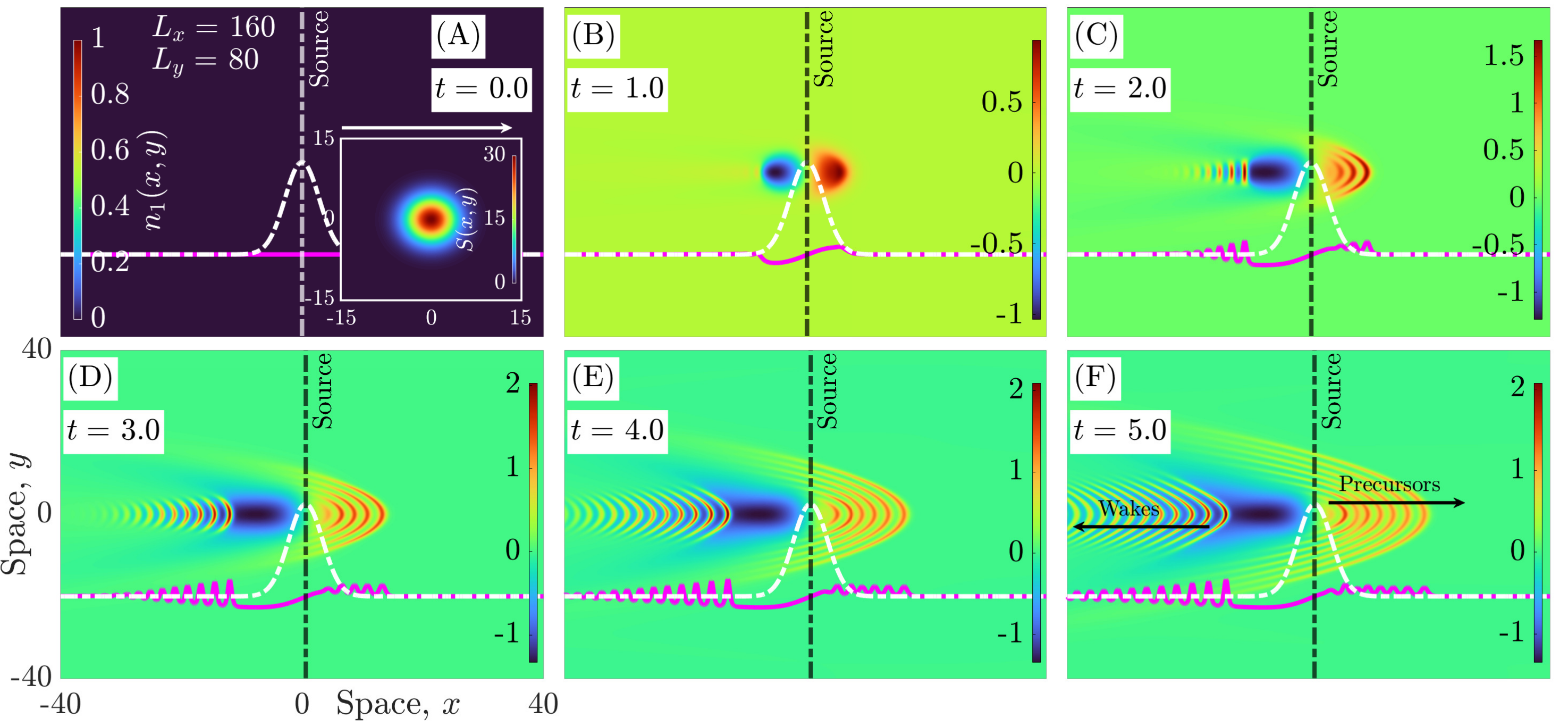}
\caption{The normalized perturbed dust density $n_1(x,y,t)$ obtained from the solution of the fKP equation (\textit{i.e.,} Eq.~\eqref{fKP_eqn}) at different time intervals showing the excitation and evolution of precursor solitons and trailing wakes. The equation coefficients have been evaluated using the experimental parameters listed in Table \ref{Tab2}. The driving source profile (shown in the inset of subplot (A)) is given by~\eqref{GaussProfile} with values of $w_x = w_y = 4.0$, $v_{dx} = 1.10$, $v_{dy} = 0.0$ and $A_s = 30$ that are taken from the experimental measurements. \textcolor{black}{All the parameters used in the figure are in normalized units.}
}
\label{Fig_1}
\end{figure*}
%%%%%%%%%%%%%%%%%%%%%%%%%%%%%%%%%%%%
%%%%%%%%%%%%%%%%%%%%%%%%%%%%%%%%%%%%%%%%%%%%%%%%%%%%%%%%%%%%%%%%%%%%%%%%%%%%%%%%
%%%%%%%%%%%%%%%%%%%%%%%%%%%%%%%%%%%%%%%%%%%%%%%%%%%%%%%%%%%%%%%%%%%%%%%%%%%%%%%%
\section{Application of the fKP model to a dusty plasma experiment}
\label{fKP_Exp}
%%%%%%%%%%%%%%%%%%%%%%%%%%%%%%%%%%%%%%%%%%%%%%%%%%%%%%%%%%%%%%%%%%%%%%%%%%%%%%%%
%%%%%%%%%%%%%%%%%%%%%%%%%%%%%%%%%%%%%%%%%%%%%%%%%%%%%%%%%%%%%%%%%%%%%%%%%%%%%%%%
We now test the capability of the fKP equation to model and reproduce actual experimental results in order to assess its practical utility. For this, we compare numerical solutions of Eq.~\eqref{fKP_equation2} obtained using parameters taken from a recent dusty plasma experiment \cite{Krishan_POP_2024} with the observed experimental results. In this experiment, a charged dust fluid was made to flow supersonically (\textit{i.e.,} with a velocity faster than the dust acoustic speed in the medium) over a stationary object. This gave rise to the excitation of regularly spaced crescent-shaped precursor solitons traveling in the fore-wake region with a velocity faster than the speed of the dust fluid. Several important features were noted that were distinct from earlier one-dimensional precursor solitons excited by a planar source \cite{Surabhi_PRE_2016}. The precursor solitons now were crescent-shaped as viewed from the top and represented half-cylindrical structures - a distinct geometrical signature of the shape of the source. Unlike the one-dimensional KdV solitons the amplitude of these curved solitons did not remain constant in time but decayed as an algebraic function of time with a scaling of $t^{-1/2}$ while their width increased as $t^{1/4}$ thus ensuring that the soliton parameter defined as the product of the amplitude with the square of the width remained a constant. 
It was further observed that the speed of the precursors increased with an increase in the velocity of the dust fluid. We now examine whether all the above-mentioned experimentally observed features can be reproduced by the fKP model both qualitatively and quantitatively.
\paragraph*{}
For this, we next present numerical solutions of the model fKP equation for the various experimental conditions. It is convenient to integrate Eq.~\eqref{fKP_equation2} with respect to $\xi$ and employ the Fourier multiplier, $\partial_{\xi}^{-1} = (i/k_{\xi})$~\cite{Ajaz_POP_2022} to express the fKP equation in the form~\cite{Klein_JNS_2007, Truitt_JSR_2021}
%%%%%%%%%%%%%%%%%%%%%%%%%%%%%%%%%%%%
\begin{equation}\label{fKP_eqn}
     \frac{\partial n_1}{\partial \tau} 
    + \alpha  n_1 \frac{\partial n_1}{\partial \xi} 
    + \beta  \frac{\partial^3 n_1}{\partial \xi^3} 
   % \\ \nonumber
    + \gamma  \partial_\xi^{-1} \frac{\partial^2 n_1}{\partial \eta^2} 
   = \delta  \frac{\partial S}{\partial \xi} , 
\end{equation}
%%%%%%%%%%%%%%%%%%%%%%%%%%%%%%%%%%%%
%\paragraph*{}
where the coefficients $\alpha = A_1 c_1$, $\beta = A_2$ determine the strength of the nonlinearity and dispersion of the medium, respectively, and $\gamma = A_3$ quantifies the effect of the transverse dimension. These quantities depend on the dusty plasma properties. $\delta = - A_2/c_1$ quantifies the contribution of the driving charged source object and depends on the charge density of the object.
 \paragraph*{}
For our numerical solution of the fKP equation Eq.~\eqref{fKP_eqn}, we have employed the pseudo-spectral method with a fourth-order Runge-Kutta (RK-4) scheme for time evolution~\cite{Klein_JNS_2007} using periodic boundary conditions in both directions.  We have also employed the Richardson extrapolation to improve the rate of convergence of the numerical solution.  The code has been validated by reproducing earlier published work based on the KP and fKP equations~\cite{Klein_JNS_2007, Truitt_JSR_2021, Acharya_PRE_2021}. To make explicit the forced excitation origin of the solitons, we set the initial perturbation of the density variable $n_1(x,y,t=0) = 0$ and launch the solutions with only a finite time-dependent source term.
%%%%%%%%%%%%%%%%%%%%%%%%%%%%%%%%%%%%%%%%%%%%%%%%%%%%%%%%%%%%%%%%%%%%%%
\begin{table}[ht!]
    \centering
    \caption{Plasma parameters used in this investigation to derive the coefficients of the fKP equation~\cite{Surabhi_RSI_2015, Krishan_POP_2023, Krishan_POP_2024}.}
{\renewcommand{\arraystretch}{1.25}
    \begin{tabular}{p{4.25cm} p{4.25cm}}
    \hline \hline
    Plasma parameter  &  Parameter value\\
    \hline \hline
      $n_{i0}$      &  $2.0 \times 10^{14}$ m$^{-3}$ \\ 
      $n_{e0}$      &  $1.0 \times 10^{14}$ m$^{-3}$ \\
      $n_{d0}$      &  $1.0 \times 10^{10}$ m$^{-3}$ \\ 
      $Z_d$         & $ 10^{4}$\\
    \hline
      $k_B T_{e}$   &  $4.000$ eV\\ 
      $k_B T_{i}$   &  $0.030$ eV \\ 
      $k_B T_{d}$   &  $0.030$ eV \\ 
%      $r_d$         & $2\ \times 10^{-6} $ m\\
      $m_d$         & $1.0\ \times 10^{-13} $ kg\\
    \hline
   \textcolor{black}{ $\lambda_D$ }    &  $128.76$ $\mu$m \\
%  \textcolor{black}{ $\lambda_D$ }     &  $83.11$  $\mu$m \\
   \textcolor{black}{ $\omega_{pd}$ }  &  $170.27$ rad/s \\
  $V_{DAW} = \lambda_D \omega_{pd}$      &  $2.19$ cm/s\\
   \hline \hline
   \textcolor{black}{fKP coefficient}  &  \textcolor{black}{Numerical value}\\
   \hline \hline
   $\alpha $ &  $+3.56$ \\
   $\beta$   &  $+0.18$ \\
   $\gamma$  &  $+0.35$ \\
   $\delta$  &  $-0.09$ \\
\hline \hline
  Floating potential & $\phi_f = 300$ V \\
  Plasma potential   & $\phi_p = 320$ V \\
  $\phi_p - \phi_f$  & $20$ V \\
  $(\phi_p - \phi_f)/T_i$  & $666.667$ \\
  Effective potential $\phi_s$   & $ \left\{ (\phi_p - \phi_f)/T_i \right\} \exp(-d/\lambda_D)$ 
  \\
% $\phi_s$  & 32 \\
% Sheath width & $d = 3\lambda_D$\\
\hline \hline
    \end{tabular}
}
    \label{Tab2}
\end{table}
%%%%%%%%%%%%%%%%%%%%%%%%%%%%%%%%%%%%%%%%%%%%%%%%%%%%%%%%%%%%%%%%%%%%%%
%\paragraph*{}
The values of the coefficients $\alpha$, $\beta$, $\gamma$, and $\delta$ are determined from the experimental plasma parameters~\cite{Krishan_POP_2024} listed in Table~\ref{Tab2}. 
\paragraph*{}
%A Gaussian form can realistically represent the potential of the moving charged source,
%interacts with the plasma through its sheath potential that 
\textcolor{black}{The potential associated with the moving charged density source can be realistically modeled using a localized Gaussian-like functional form that propagates in space at a constant speed and is given by}
%%%%%%%%%%%%%%%%%%%%%%%%%%%%%%%%%%%%
\begin{equation}\label{GaussProfile}
 S = A_s \exp\left\{
 - \left(\frac{x - x_0 + F_{x} t}{w_x}\right)^2  
 - \left(\frac{y - y_0 + F_{y} t}{w_y}\right)^2
 \right\} ,
\end{equation}
%%%%%%%%%%%%%%%%%%%%%%%%%%%%%%%%%%%%
where $F_x = 1 - v_{dx}$ and $F_y = 0$.  
\textcolor{black}{This parameterization captures the essential features of the experimental forcing mechanism described in Ref.~\cite{Krishan_POP_2024}, where the charged object interacts with the plasma through its Debye shielded floating potential that physically has a Gaussian form. Thus, our parametrization conforms to the boundary conditions of the experiment, both close to the source and far away from it.  The charge distribution on the source remains localized in the frame of the dust fluid medium, consistent with the finite size of the experimental driver. Finally, the velocity of the source, a key parameter for the generation of precursor solitons, is taken to exactly match the experimental value of the flow velocity given to the dusty plasma medium.}
%\paragraph*{}
%%%%%%%%%%%%%%%%%%%%%%%%%%%%%%%%%%%%
\begin{figure} [h!]
\includegraphics[width=\columnwidth]{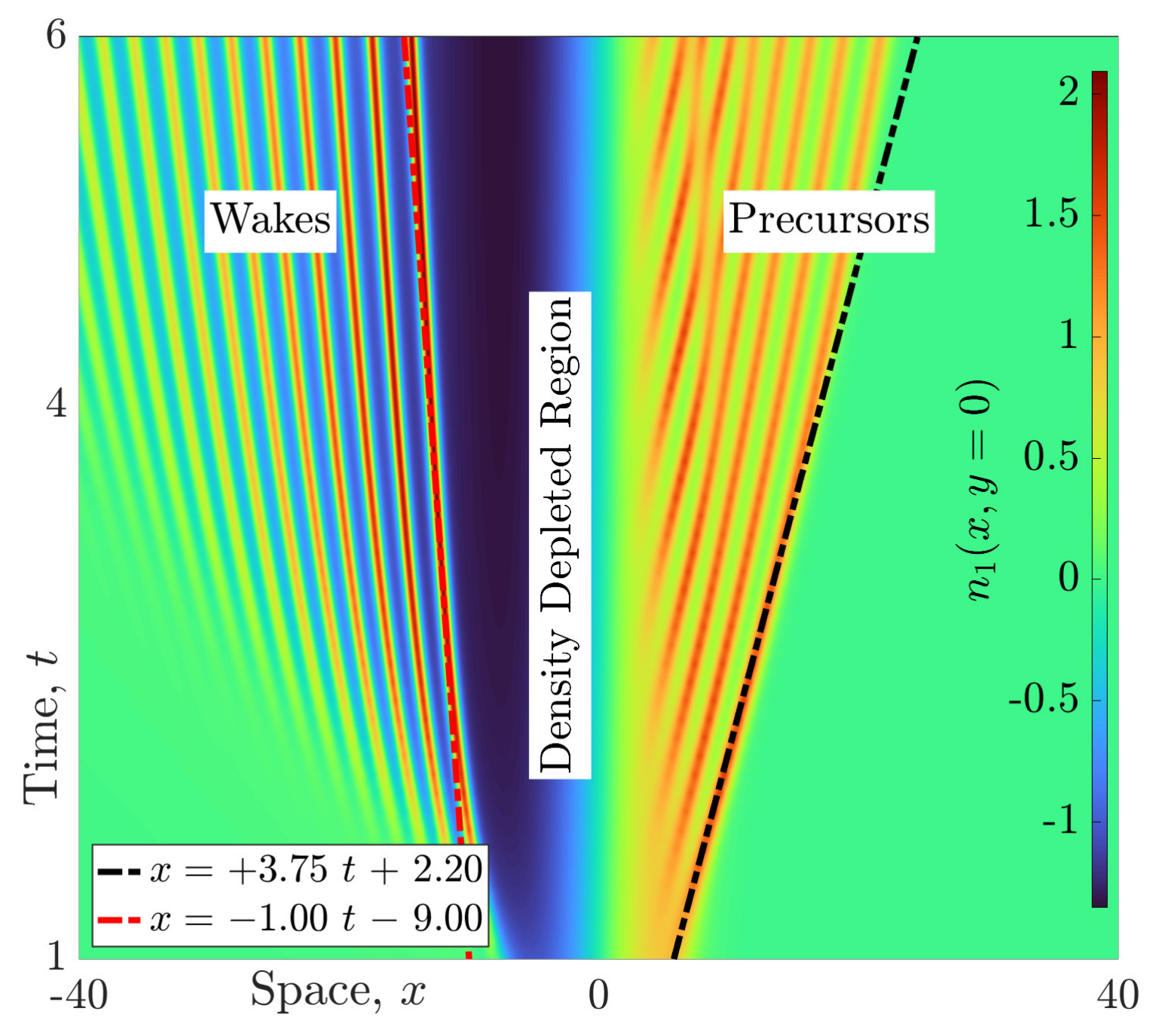}
%\includesvg[width = \columnwidth]{figure_2.svg}
\caption{
Space-Time evolution plot of the precursor solitons and wakes shown in Fig.~\ref{Fig_1}. The slope of the black dashed line provides the velocity of the precursor. \textcolor{black}{All the parameters used in the figures are in normalized units.} }
\label{Fig_2}
\end{figure}
%%%%%%%%%%%%%%%%%%%%%%%%%%%%%%%%%%%%
\paragraph*{}
We consider one of the experimental runs where the source is taken to move in the positive x-direction with a velocity $2.40$ cm/s, which in normalized units gives $v_{dx} = 1.10$ and its y-component of velocity $v_{dy} = 0$. To determine the value of the amplitude $A_s$ we consider the experimentally estimated value of the source potential, given by the difference of the floating potential and the plasma potential, as $\phi = 20.0\ V$. In the normalized units employed in the model equation, this would be $e\phi / T_i$, which for an ion temperature of $0.03$ eV gives $\phi = 666.667$. This is reduced to $\phi e^{-d/\lambda_D}$ at a distance $d$ from the source surface - the region where the precursors originate. For a stationary source (or a non-flowing plasma), $d$ would be of the order of the Debye shielding distance, typically a few times the Debye length. For a flowing plasma, this distance gets reduced as the plasma particles are able to penetrate deeper into the Debye sphere and get closer to the source surface. For the experimental velocity of $2.40$ cm/s, taking  $d \sim 2.45 \lambda_D$ gives us a value of $A_s = 30$. We further choose the values of $w_x = w_y = 4$ to remain close to the experimental values. 
%%%%%%%%%%%%%%%%%%%%%%%%%%%%%%%%%%%%%%%%%%%%%%%%%%%%%%%%%%%%%%%%%%%%%%%%%%
\paragraph*{}
The numerical solution of the fKP equation (\textit{i.e.,}~Eq.~\eqref{fKP_eqn}), for the above-mentioned parameters, is shown in Fig.~\ref{Fig_1}. It can be seen that the moving source produces a succession of crescent-shaped precursors, thus faithfully reproducing the geometric feature of the cylindrical solitons observed in the experiment~\cite{Krishan_POP_2024}. 
%%%%%%%%%%%%%%%%%%%%%%%%%%%%%%%%%%%%
\begin{figure*} [ht!]
\includegraphics[width = \textwidth]{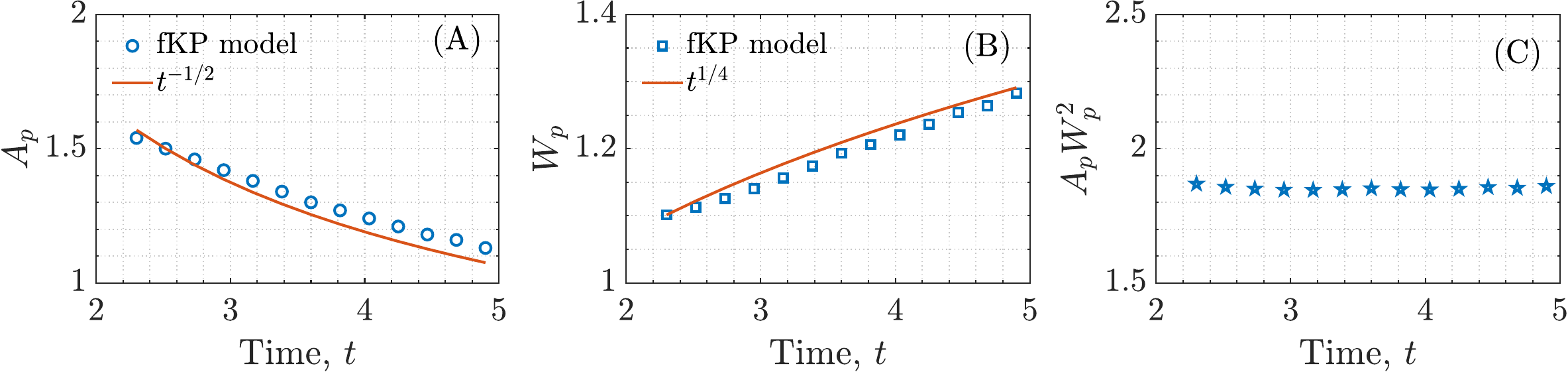}
%\includesvg[width = \textwidth]{Figure_1.svg}
\caption{(A) The nature of the decay of the amplitude $A_p$ of the two-dimensional precursor soliton and (B) the broadening of its width $W_p$ as a function of time with characteristic scaling of $t^{-1/2}$ and $t^{1/4}$, respectively. The subplot (C) shows the constancy of the soliton parameter $A_p W_p^{2}$.
}
\label{Fig_3}
\end{figure*}
%%%%%%%%%%%%%%%%%%%%%%%%%%%%%%%%%%%%
%%%%%%%%%%%%%%%%%%%%%%%%%%%%%%%%%%%%
\begin{figure} [ht!]
\includegraphics[width=\columnwidth]{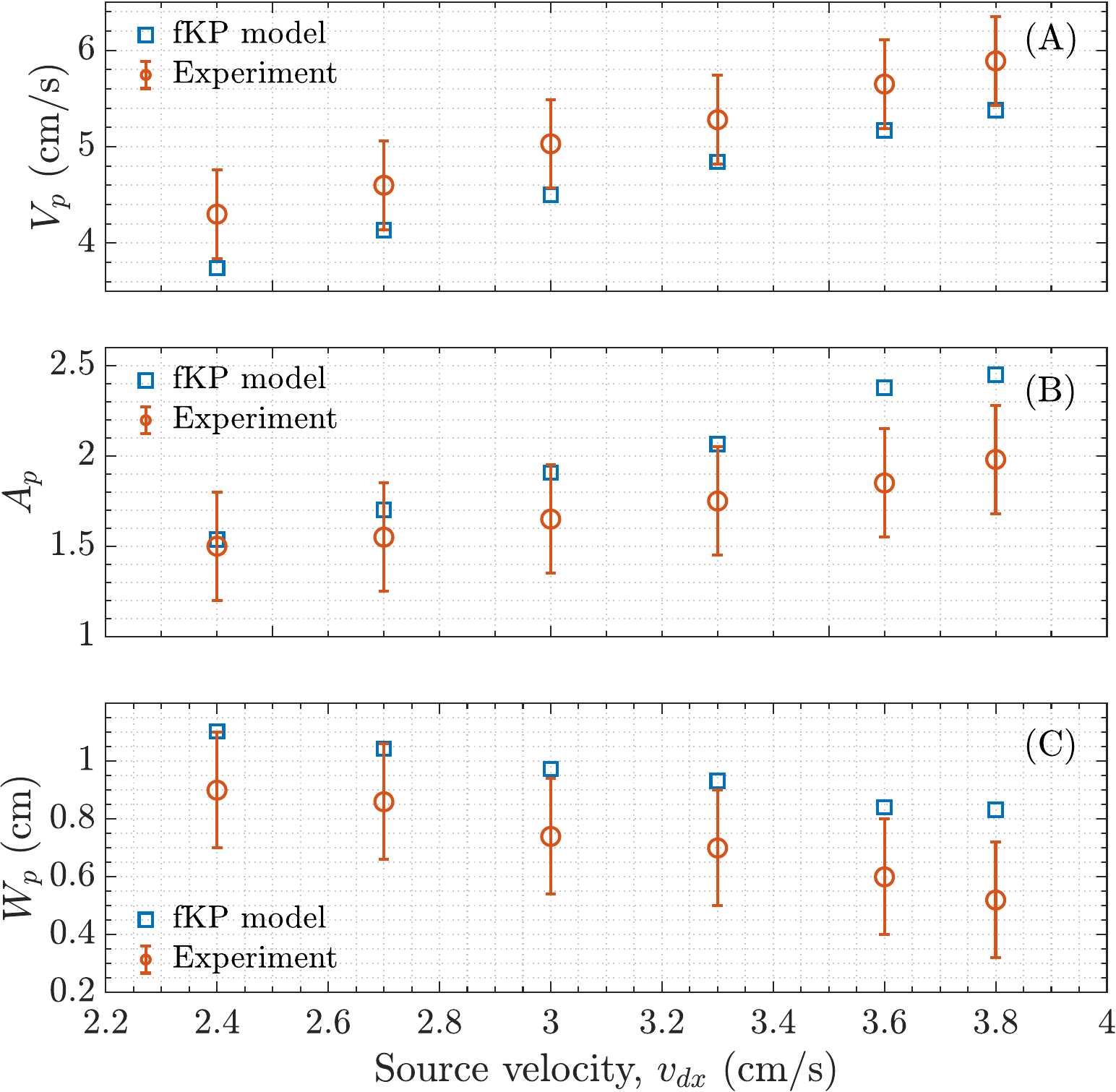}
%\includesvg[width = \columnwidth]{Figure_2.svg}
\caption{A comparison of the experimental~\cite{Krishan_POP_2024} and fKP model solution values of the velocity $(V_p)$, amplitude $(A_p)$, and width $(W_p)$ of the precursor solitons as a function of the source velocity $v_{dx}$.}
\label{Fig_4}
\end{figure}
%%%%%%%%%%%%%%%%%%%%%%%%%%%%%%%%%%%%
\textcolor{black}{Figure~\ref{Fig_1} represents the spatio-temporal evolution of the normalized 2D perturbed density $n_1(x,y,t)$ of the dusty plasma medium propagating over the charged object, which is modeled by the fKP equation. The density perturbations are excited by the localized, 2D space-time-dependent charge source, propagating in the positive x-direction. The subplot (A) represents the initial perturbed density of the dusty plasma medium. The inset in subplot (A) depicts the localized, space-time-dependent Gaussian source (\textit{i.e.,} source charge density). The subplots (C)-(F) show the evolution of precursor solitons (density-enhanced structures), which are propagating ahead of the supersonically streaming charged source. These crescent-like structures resemble qualitatively the nonlinear structures that were observed in the dusty plasma medium~\cite{Krishan_POP_2024}. The structures that are traveling behind the source are the wakes. The location of the source in each subplot is shown by the vertical black dashed line.}
\paragraph*{}
\textcolor{black}{Figure~\ref{Fig_2}, displays the temporal evolution of normalized perturbed density $n_1(x,y,t)$, obtained by collecting time-series data at one arbitrary spatial location, from the numerical solution of the fKP equation. With the advancement of time, the charged source excites more and more precursor solitons ahead of it, which move faster than the source. These precursor structures with enhanced dust densities (yellow color) travel along x at a constant pace with a supersonic velocity of $V_p = 3.75$. The velocity of these nonlinear structures can be deduced from the straight line fit (black dashed slant line) of $x = 3.75t + 2.20$ to the trajectory of the precursor. Figure~\ref{Fig_2} also shows that with time, the density-depleted region keeps on expanding. This is because the source generates more and more density-enhanced structures (precursor solitons) ahead of it, and mass is transferred from the region behind the source to conserve the mass. The dispersive structures, which are trailing behind the density-depleted region, are the wake structures.} 
%The velocity of a given soliton can be estimated from the slope of its space-time plot as shown in Fig.~\ref{Fig_2}. 
In this particular case, the velocity of the precursor soliton is $V_p=3.75$, which is quite close to the experimental value. The numerically measured values of the amplitude and width (full width at half maximum) of the precursor are found to be $A_p=1.54$ and $W_p=1.10$, respectively, again closely matching the experimental results. We have next examined the temporal evolution of the amplitude (Fig.~\ref{Fig_3}(A)) and width (Fig.~\ref{Fig_3}(B)) of a given precursor solution and found them to decay and grow, respectively, with scaling of $t^{-1/2}$ and $t^{1/4}$. \textcolor{black}{However, the precursor solitons obey the constancy of $A_pW_p^2$ (Fig.~\ref{Fig_3}(C)), the hallmark of solitons, during their evolution in time.} 
%%%%%%%%%%%%%%%%%%%%%%%%%%%%%%%%%%%%
\begin{figure}[h!]
\includegraphics[width=\columnwidth]{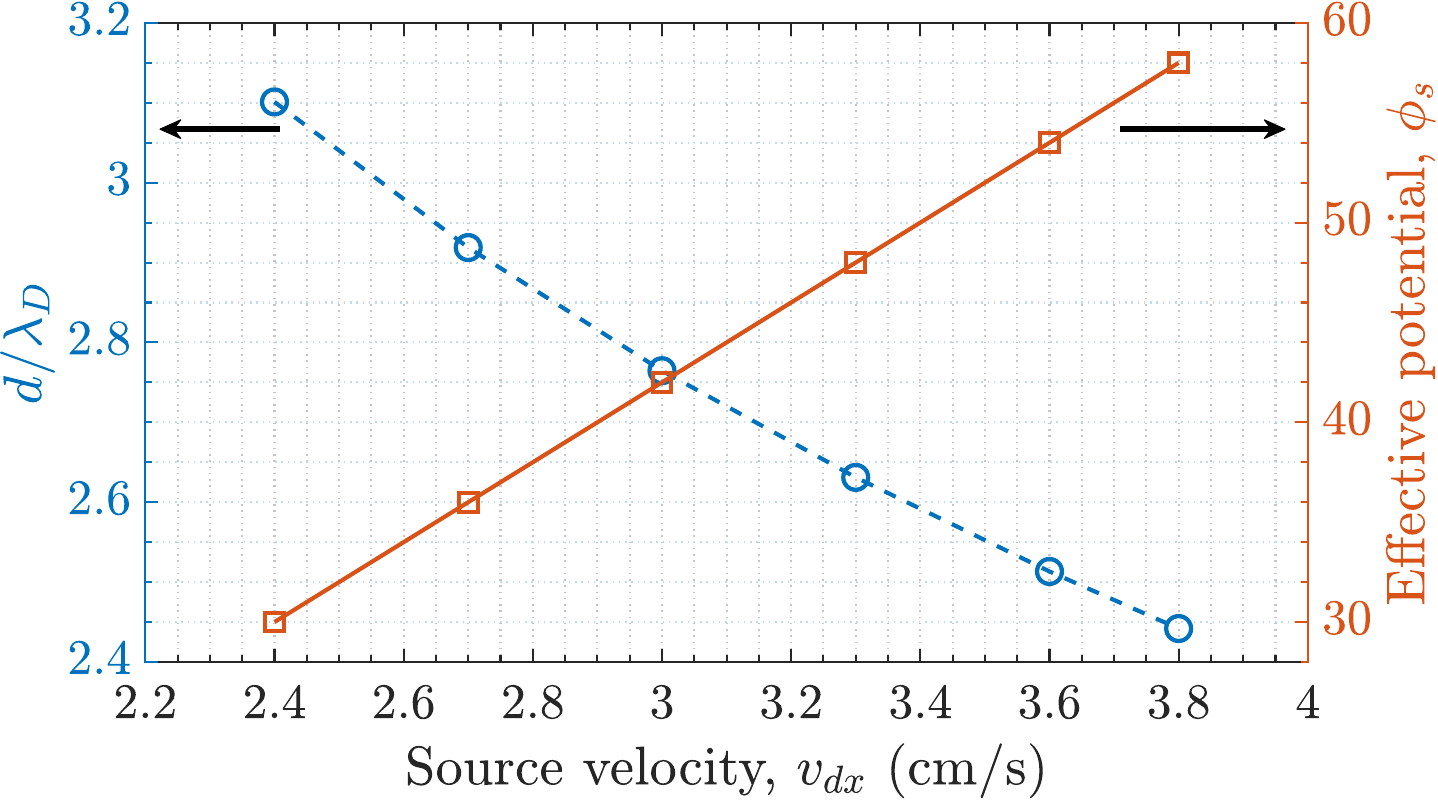}
%\includesvg[width = \columnwidth]{Figure_2.svg}
\caption{Variation of effective potential $(\phi_s)$ and $(d/\lambda_D)$ as a function of the source velocity $v_{dx}$. 
}
\label{Fig_5}
\end{figure}
%%%%%%%%%%%%%%%%%%%%%%%%%%%%%%%%%%%% 
\paragraph*{}
The above procedure is repeated for all the other experimental values by keeping the same coefficients of the fKP equation but varying the values of the source velocity and the effective amplitude of source potential $A_s$ (see Fig.~\ref{Fig_4}).
\textcolor{black}{Figure~\ref{Fig_4}, captures the velocity, amplitude, and width of the precursor solitons obtained from the fKP model, as a function of the velocity of the source charge density. Figures~\ref{Fig_4}(A) and~\ref {Fig_4}(B) show the increase in the velocity and amplitude of the precursor solitons, with an increase in the velocity of the propagating source charge density in the dusty plasma medium. However, in Fig.~\ref{Fig_4}(C), there is a decrease in width with an increase in the velocity of the source charge density in the dusty plasma medium. 
%The constancy of the quantity $A_pW_p^2$ holds for the precursor solitons.
We have also plotted the values of the corresponding physical quantities of the precursor solitons obtained from the laboratory-based dusty plasma experiment~\cite{Krishan_POP_2024} for a one-to-one comparison.} Furthermore, $A_s$ is varied by decreasing the value of $d$ in a nearly linear fashion as a function of the source velocity to account for the fact that at higher flow velocities the region of interaction becomes closer to the source surface (see Fig.~\ref{Fig_5}). Using such a scaling, we have found an excellent agreement between the amplitudes of the fKP precursors and the observed experimental results (which have an error bar range of $10-20\%$) for the various values of the source velocity, as shown in Fig.~\ref{Fig_4}.
Thus, the fKP model reproduces all the experimental characteristics of the dusty plasma experiment~\cite{Krishan_POP_2024} with a close quantitative agreement with the measured experimental quantities. The velocity of the precursor solitons obtained from the fKP model and the dusty plasma experiment~\cite{Krishan_POP_2024} are tabulated in Table~\ref{Tab3}.
%%%%%%%%%%%%%%%%%%%%%%%%%%%%%%%%%%%%%%%%%%%%%%%%%%%%%%%%%%%%%%%%%%%%%%
\begin{table}[h!]
    \centering
    \caption{Comparison of precursor soliton velocity for the dusty plasma experiment~\cite{Krishan_POP_2024} and the fKP simulation.}
{\renewcommand{\arraystretch}{1.25}
    \begin{tabular}{p{3.5cm} p{5.0cm}}
    \hline \hline
    Source velocity (cm/s)  & \hspace{0.50cm} Soliton velocity  (cm/s)      \\
                     & Experiment~\cite{Krishan_POP_2024}   \hspace{0.50cm}   fKP model \\
    \hline \hline
      $2.40$      &  $4.30 \pm 0.45$ \hspace{1.50cm} $3.74$ \\ 
      $2.70$      &  $4.60 \pm  0.47$ \hspace{1.50cm} $4.13$ \\
      $3.00$      &  $5.03 \pm  0.54$ \hspace{1.50cm} $4.49$ \\
      $3.30$      &  $5.28 \pm  0.42$ \hspace{1.50cm} $4.84$ \\
      $3.60$      &  $5.65 \pm  0.48$ \hspace{1.50cm} $5.17$ \\
      $3.80$      &  $5.89 \pm  0.49$ \hspace{1.50cm} $5.38$ \\
    \hline
    \end{tabular}
}
    \label{Tab3}
\end{table}
%%%%%%%%%%%%%%%%%%%%%%%%%%%%%%%%%%%%%%%%%%%%%%%%%%%%%%%%%%%%%%%%%%%%%%
%%%%%%%%%%%%%%%%%%%%%%%%%%%%%%%%%%%%%%%%%%%%%%%%%%%%%%%%%%%%%%%%%%%%%%%%%%%%%%%%
%%%%%%%%%%%%%%%%%%%%%%%%%%%%%%%%%%%%%%%%%%%%%%%%%%%%%%%%%%%%%%%%%%%%%%%%%%%%%%%%%
\section{Summary and conclusion}
\label{sum_con}
%%%%%%%%%%%%%%%%%%%%%%%%%%%%%%%%%%%%%%%%%%%%%%%%%%%%%%%%%%%%%%%%%%%%%%%%%%%%%%%%
%%%%%%%%%%%%%%%%%%%%%%%%%%%%%%%%%%%%%%%%%%%%%%%%%%%%%%%%%%%%%%%%%%%%%%%%%%%%%%%%%
\textcolor{black}{
To summarise, in our present work, we have examined the validity and applicability of the forced Kadomtsev-Petviashvili equation to realistically model the excitation and dynamical evolution of precursor solitons in a plasma. This has been done by comparing the numerical solutions of the equation with experimental findings of a dusty plasma experiment, where such solitons were excited by making a dusty plasma flow over a charged cylindrical object. For this, the model fKP equation appropriate for a dusty plasma medium has been derived, and its coefficient values have been appropriately fixed from the experimental parameters. Our numerical solutions obtained for the various values of the source velocities used in the experiment are then compared with the experimental findings. It is seen that the numerical solutions reproduce quite well the geometrical features of the experimental solitons. They also closely match the values of their velocities, amplitudes, and widths to within the range of experimental errors.} The residual differences can arise from several factors, such as the lack of damping effects in the model equation and the errors associated with determining the various coefficients of the model equation. It should be mentioned that the numerical results are quite sensitive to the value of the effective source potential that drives the fluctuations to grow and evolve into a precursor soliton. The distance $d$ from the source surface at which this happens is not a constant but varies with the source velocity (or the plasma flow velocity in the frame of the source). We find that a near-linear decrease of $d$ with the source velocity, as shown in Fig.~\ref{Fig_5}, provides an excellent match to the experimental results. The decrease in $d$ can be physically understood from the fact that for a higher flow velocity, there is a larger accumulation of plasma closer to the surface that can support the fluctuations. An experimental verification of such scaling can be easily carried out by fluctuation measurements close to the source surface for different plasma flow velocities and would help provide useful insights into the birth mechanism of precursor solitons. 
\paragraph*{}
To conclude, our present findings based on the first detailed comparison of the fKP model against experimental results provide a useful benchmark of the validity and utility of the model for describing precursor solitonic emissions that may help promote its use in practical applications like debris detection. It can serve as a basis for further refinements and generalizations of the model to include effects like collisional damping, background viscosity, \textit{etc.}~\cite{Ajaz_PRE_2023, Ott_POF_1969} that have been ignored in our present work. It can also inspire new lines of experimental work to explore the dynamics of the plasma close to the charged source.
%%%%%%%%%%%%%%%%%%%%%%%%%%%%%%%%%%%%%%%%%%%%%%%%%%%%%%%%%%%%%%%%%%%%%%%%%%%%%%%%
%%%%%%%%%%%%%%%%%%%%%%%%%%%%%%%%%%%%%%%%%%%%%%%%%%%%%%%%%%%%%%%%%%%%%%%%%%%%%%%%%
\begin{acknowledgements}
A. S. acknowledges the support provided for this research by the Asian Office of Aerospace Research and Development (AOARD) under Grant No. FA9550-23-1-0003 and is thankful to the Indian National Science Academy (INSA) for the INSA Honorary Scientist position.
\end{acknowledgements}
%%%%%%%%%%%%%%%%%%%%%%%%%%%%%%%%%%%%%%%%%%%%%%%%%%%%%%%%%%%%%%%%%%%%%%%%%%%%%%%%
%%%%%%%%%%%%%%%%%%%%%%%%%%%%%%%%%%%%%%%%%%%%%%%%%%%%%%%%%%%%%%%%%%%%%%%%%%%%%%%%
\bibliography{PrecursorPlasma}

\begin{thebibliography}{36}%
\makeatletter
\providecommand \@ifxundefined [1]{%
 \@ifx{#1\undefined}
}%
\providecommand \@ifnum [1]{%
 \ifnum #1\expandafter \@firstoftwo
 \else \expandafter \@secondoftwo
 \fi
}%
\providecommand \@ifx [1]{%
 \ifx #1\expandafter \@firstoftwo
 \else \expandafter \@secondoftwo
 \fi
}%
\providecommand \natexlab [1]{#1}%
\providecommand \enquote  [1]{``#1''}%
\providecommand \bibnamefont  [1]{#1}%
\providecommand \bibfnamefont [1]{#1}%
\providecommand \citenamefont [1]{#1}%
\providecommand \href@noop [0]{\@secondoftwo}%
\providecommand \href [0]{\begingroup \@sanitize@url \@href}%
\providecommand \@href[1]{\@@startlink{#1}\@@href}%
\providecommand \@@href[1]{\endgroup#1\@@endlink}%
\providecommand \@sanitize@url [0]{\catcode `\\12\catcode `\$12\catcode `\&12\catcode `\#12\catcode `\^12\catcode `\_12\catcode `\%12\relax}%
\providecommand \@@startlink[1]{}%
\providecommand \@@endlink[0]{}%
\providecommand \url  [0]{\begingroup\@sanitize@url \@url }%
\providecommand \@url [1]{\endgroup\@href {#1}{\urlprefix }}%
\providecommand \urlprefix  [0]{URL }%
\providecommand \Eprint [0]{\href }%
\providecommand \doibase [0]{https://doi.org/}%
\providecommand \selectlanguage [0]{\@gobble}%
\providecommand \bibinfo  [0]{\@secondoftwo}%
\providecommand \bibfield  [0]{\@secondoftwo}%
\providecommand \translation [1]{[#1]}%
\providecommand \BibitemOpen [0]{}%
\providecommand \bibitemStop [0]{}%
\providecommand \bibitemNoStop [0]{.\EOS\space}%
\providecommand \EOS [0]{\spacefactor3000\relax}%
\providecommand \BibitemShut  [1]{\csname bibitem#1\endcsname}%
\let\auto@bib@innerbib\@empty
%</preamble>
\bibitem [{\citenamefont {Sen}\ \emph {et~al.}(2015)\citenamefont {Sen}, \citenamefont {Tiwari}, \citenamefont {Mishra},\ and\ \citenamefont {Kaw}}]{Sen_ASR_2015}%
  \BibitemOpen
  \bibfield  {author} {\bibinfo {author} {\bibfnamefont {A.}~\bibnamefont {Sen}}, \bibinfo {author} {\bibfnamefont {S.}~\bibnamefont {Tiwari}}, \bibinfo {author} {\bibfnamefont {S.}~\bibnamefont {Mishra}},\ and\ \bibinfo {author} {\bibfnamefont {P.}~\bibnamefont {Kaw}},\ }\bibfield  {title} {\bibinfo {title} {Nonlinear wave excitations by orbiting charged space debris objects},\ }\href {https://doi.org/https://doi.org/10.1016/j.asr.2015.03.021} {\bibfield  {journal} {\bibinfo  {journal} {Adv. Space Res.}\ }\textbf {\bibinfo {volume} {56}},\ \bibinfo {pages} {429 } (\bibinfo {year} {2015})}\BibitemShut {NoStop}%
\bibitem [{\citenamefont {Truitt}\ and\ \citenamefont {Hartzell}(2020{\natexlab{a}})}]{Truitt_JSR_2020A}%
  \BibitemOpen
  \bibfield  {author} {\bibinfo {author} {\bibfnamefont {A.~S.}\ \bibnamefont {Truitt}}\ and\ \bibinfo {author} {\bibfnamefont {C.~M.}\ \bibnamefont {Hartzell}},\ }\bibfield  {title} {\bibinfo {title} {{Simulating Plasma Solitons from Orbital Debris Using the Forced Korteweg–de Vries Equation}},\ }\href {https://doi.org/10.2514/1.A34652} {\bibfield  {journal} {\bibinfo  {journal} {J. Spacecraft Rockets}\ }\textbf {\bibinfo {volume} {57}},\ \bibinfo {pages} {876} (\bibinfo {year} {2020}{\natexlab{a}})}\BibitemShut {NoStop}%
\bibitem [{\citenamefont {Truitt}\ and\ \citenamefont {Hartzell}(2020{\natexlab{b}})}]{Truitt_JSR_2020B}%
  \BibitemOpen
  \bibfield  {author} {\bibinfo {author} {\bibfnamefont {A.~S.}\ \bibnamefont {Truitt}}\ and\ \bibinfo {author} {\bibfnamefont {C.~M.}\ \bibnamefont {Hartzell}},\ }\bibfield  {title} {\bibinfo {title} {Simulating damped ion acoustic solitary waves from orbital debris},\ }\href {https://doi.org/10.2514/1.A34674} {\bibfield  {journal} {\bibinfo  {journal} {J. Spacecr. Rockets}\ }\textbf {\bibinfo {volume} {57}},\ \bibinfo {pages} {975} (\bibinfo {year} {2020}{\natexlab{b}})}\BibitemShut {NoStop}%
\bibitem [{\citenamefont {Truitt}\ and\ \citenamefont {Hartzell}(2021)}]{Truitt_JSR_2021}%
  \BibitemOpen
  \bibfield  {author} {\bibinfo {author} {\bibfnamefont {A.~S.}\ \bibnamefont {Truitt}}\ and\ \bibinfo {author} {\bibfnamefont {C.~M.}\ \bibnamefont {Hartzell}},\ }\bibfield  {title} {\bibinfo {title} {{Three-Dimensional Kadomtsev–Petviashvili Damped Forced Ion Acoustic Solitary Waves from Orbital Debris}},\ }\href {https://doi.org/10.2514/1.A34805} {\bibfield  {journal} {\bibinfo  {journal} {J. Spacecr. Rockets}\ }\textbf {\bibinfo {volume} {58}},\ \bibinfo {pages} {848} (\bibinfo {year} {2021})}\BibitemShut {NoStop}%
\bibitem [{\citenamefont {Acharya}\ \emph {et~al.}(2021)\citenamefont {Acharya}, \citenamefont {Mukherjee},\ and\ \citenamefont {Janaki}}]{Acharya_PRE_2021}%
  \BibitemOpen
  \bibfield  {author} {\bibinfo {author} {\bibfnamefont {S.~P.}\ \bibnamefont {Acharya}}, \bibinfo {author} {\bibfnamefont {A.}~\bibnamefont {Mukherjee}},\ and\ \bibinfo {author} {\bibfnamefont {M.~S.}\ \bibnamefont {Janaki}},\ }\bibfield  {title} {\bibinfo {title} {Bending of pinned dust-ion acoustic solitary waves in presence of charged space debris},\ }\href {https://doi.org/10.1103/PhysRevE.104.014214} {\bibfield  {journal} {\bibinfo  {journal} {Phys. Rev. E}\ }\textbf {\bibinfo {volume} {104}},\ \bibinfo {pages} {014214} (\bibinfo {year} {2021})}\BibitemShut {NoStop}%
\bibitem [{\citenamefont {Sen}\ \emph {et~al.}(2023)\citenamefont {Sen}, \citenamefont {Mukherjee}, \citenamefont {Yadav}, \citenamefont {Crabtree},\ and\ \citenamefont {Ganguli}}]{Sen_POP_2023}%
  \BibitemOpen
  \bibfield  {author} {\bibinfo {author} {\bibfnamefont {A.}~\bibnamefont {Sen}}, \bibinfo {author} {\bibfnamefont {R.}~\bibnamefont {Mukherjee}}, \bibinfo {author} {\bibfnamefont {S.~K.}\ \bibnamefont {Yadav}}, \bibinfo {author} {\bibfnamefont {C.}~\bibnamefont {Crabtree}},\ and\ \bibinfo {author} {\bibfnamefont {G.}~\bibnamefont {Ganguli}},\ }\bibfield  {title} {\bibinfo {title} {{Electromagnetic pinned solitons for space debris detection}},\ }\href {https://doi.org/10.1063/5.0099201} {\bibfield  {journal} {\bibinfo  {journal} {Phys. Plasmas}\ }\textbf {\bibinfo {volume} {30}},\ \bibinfo {pages} {012301} (\bibinfo {year} {2023})}\BibitemShut {NoStop}%
\bibitem [{\citenamefont {Dharodi}\ \emph {et~al.}(2023)\citenamefont {Dharodi}, \citenamefont {Kumar},\ and\ \citenamefont {Sen}}]{Vikram_PRE_2023}%
  \BibitemOpen
  \bibfield  {author} {\bibinfo {author} {\bibfnamefont {V.}~\bibnamefont {Dharodi}}, \bibinfo {author} {\bibfnamefont {A.}~\bibnamefont {Kumar}},\ and\ \bibinfo {author} {\bibfnamefont {A.}~\bibnamefont {Sen}},\ }\bibfield  {title} {\bibinfo {title} {Signatures of an energetic charged body streaming in a plasma},\ }\href {https://doi.org/10.1103/PhysRevE.107.025207} {\bibfield  {journal} {\bibinfo  {journal} {Phys. Rev. E}\ }\textbf {\bibinfo {volume} {107}},\ \bibinfo {pages} {025207} (\bibinfo {year} {2023})}\BibitemShut {NoStop}%
\bibitem [{\citenamefont {Bernhardt}\ \emph {et~al.}(2023)\citenamefont {Bernhardt}, \citenamefont {Scott}, \citenamefont {Howarth},\ and\ \citenamefont {Morales}}]{Bernhardt_POP_2023}%
  \BibitemOpen
  \bibfield  {author} {\bibinfo {author} {\bibfnamefont {P.~A.}\ \bibnamefont {Bernhardt}}, \bibinfo {author} {\bibfnamefont {L.}~\bibnamefont {Scott}}, \bibinfo {author} {\bibfnamefont {A.}~\bibnamefont {Howarth}},\ and\ \bibinfo {author} {\bibfnamefont {G.~J.}\ \bibnamefont {Morales}},\ }\bibfield  {title} {\bibinfo {title} {{Observations of plasma waves generated by charged space objects}},\ }\href {https://doi.org/10.1063/5.0155454} {\bibfield  {journal} {\bibinfo  {journal} {Phys. Plasmas}\ }\textbf {\bibinfo {volume} {30}},\ \bibinfo {pages} {092106} (\bibinfo {year} {2023})}\BibitemShut {NoStop}%
\bibitem [{\citenamefont {Lai}(2012)}]{Lai_IEEE_2012}%
  \BibitemOpen
  \bibfield  {author} {\bibinfo {author} {\bibfnamefont {S.~T.}\ \bibnamefont {Lai}},\ }\href {https://doi.org/http://ieeexplore.ieee.org/document/9452389} {\emph {\bibinfo {title} {{Fundamentals of Spacecraft Charging: Spacecraft Interactions with Space Plasmas}}}}\ (\bibinfo  {publisher} {Princeton University Press},\ \bibinfo {year} {2012})\BibitemShut {NoStop}%
\bibitem [{\citenamefont {Tiwari}\ \emph {et~al.}(2024)\citenamefont {Tiwari}, \citenamefont {Sharma}, \citenamefont {Mishra},\ and\ \citenamefont {Sen}}]{Sanat_AA_2024}%
  \BibitemOpen
  \bibfield  {author} {\bibinfo {author} {\bibfnamefont {S.~K.}\ \bibnamefont {Tiwari}}, \bibinfo {author} {\bibfnamefont {S.}~\bibnamefont {Sharma}}, \bibinfo {author} {\bibfnamefont {S.}~\bibnamefont {Mishra}},\ and\ \bibinfo {author} {\bibfnamefont {A.}~\bibnamefont {Sen}},\ }\bibfield  {title} {\bibinfo {title} {{Charging of space debris in the LEO and GEO regions}},\ }\href {https://doi.org/https://doi.org/10.1016/j.actaastro.2024.06.008} {\bibfield  {journal} {\bibinfo  {journal} {Acta Astronautica}\ }\textbf {\bibinfo {volume} {222}},\ \bibinfo {pages} {156} (\bibinfo {year} {2024})}\BibitemShut {NoStop}%
\bibitem [{\citenamefont {Jaiswal}\ \emph {et~al.}(2016)\citenamefont {Jaiswal}, \citenamefont {Bandyopadhyay},\ and\ \citenamefont {Sen}}]{Surabhi_PRE_2016}%
  \BibitemOpen
  \bibfield  {author} {\bibinfo {author} {\bibfnamefont {S.}~\bibnamefont {Jaiswal}}, \bibinfo {author} {\bibfnamefont {P.}~\bibnamefont {Bandyopadhyay}},\ and\ \bibinfo {author} {\bibfnamefont {A.}~\bibnamefont {Sen}},\ }\bibfield  {title} {\bibinfo {title} {Experimental observation of precursor solitons in a flowing complex plasma},\ }\href {https://doi.org/10.1103/PhysRevE.93.041201} {\bibfield  {journal} {\bibinfo  {journal} {Phys. Rev. E}\ }\textbf {\bibinfo {volume} {93}},\ \bibinfo {pages} {041201} (\bibinfo {year} {2016})}\BibitemShut {NoStop}%
\bibitem [{\citenamefont {Wu}(1987)}]{Wu_JFM_1987}%
  \BibitemOpen
  \bibfield  {author} {\bibinfo {author} {\bibfnamefont {T.~Y.-T.}\ \bibnamefont {Wu}},\ }\bibfield  {title} {\bibinfo {title} {Generation of upstream advancing solitons by moving disturbances},\ }\href {https://doi.org/10.1017/S0022112087002817} {\bibfield  {journal} {\bibinfo  {journal} {J. Fluid Mech.}\ }\textbf {\bibinfo {volume} {184}},\ \bibinfo {pages} {75–99} (\bibinfo {year} {1987})}\BibitemShut {NoStop}%
\bibitem [{\citenamefont {Lee}\ \emph {et~al.}(1989)\citenamefont {Lee}, \citenamefont {Yates},\ and\ \citenamefont {Wu}}]{Lee_JFM_1989}%
  \BibitemOpen
  \bibfield  {author} {\bibinfo {author} {\bibfnamefont {S.-J.}\ \bibnamefont {Lee}}, \bibinfo {author} {\bibfnamefont {G.~T.}\ \bibnamefont {Yates}},\ and\ \bibinfo {author} {\bibfnamefont {T.~Y.}\ \bibnamefont {Wu}},\ }\bibfield  {title} {\bibinfo {title} {Experiments and analyses of upstream-advancing solitary waves generated by moving disturbances},\ }\href {https://doi.org/10.1017/S0022112089000492} {\bibfield  {journal} {\bibinfo  {journal} {J. Fluid Mech.}\ }\textbf {\bibinfo {volume} {199}},\ \bibinfo {pages} {569–593} (\bibinfo {year} {1989})}\BibitemShut {NoStop}%
\bibitem [{\citenamefont {Barkan}\ \emph {et~al.}(1994)\citenamefont {Barkan}, \citenamefont {D'Angelo},\ and\ \citenamefont {Merlino}}]{Barkan_PRL_1994}%
  \BibitemOpen
  \bibfield  {author} {\bibinfo {author} {\bibfnamefont {A.}~\bibnamefont {Barkan}}, \bibinfo {author} {\bibfnamefont {N.}~\bibnamefont {D'Angelo}},\ and\ \bibinfo {author} {\bibfnamefont {R.~L.}\ \bibnamefont {Merlino}},\ }\bibfield  {title} {\bibinfo {title} {{Charging of Dust Grains in a Plasma}},\ }\href {https://doi.org/10.1103/PhysRevLett.73.3093} {\bibfield  {journal} {\bibinfo  {journal} {Phys. Rev. Lett.}\ }\textbf {\bibinfo {volume} {73}},\ \bibinfo {pages} {3093} (\bibinfo {year} {1994})}\BibitemShut {NoStop}%
\bibitem [{\citenamefont {Jaiswal}\ \emph {et~al.}(2015)\citenamefont {Jaiswal}, \citenamefont {Bandyopadhyay},\ and\ \citenamefont {Sen}}]{Surabhi_RSI_2015}%
  \BibitemOpen
  \bibfield  {author} {\bibinfo {author} {\bibfnamefont {S.}~\bibnamefont {Jaiswal}}, \bibinfo {author} {\bibfnamefont {P.}~\bibnamefont {Bandyopadhyay}},\ and\ \bibinfo {author} {\bibfnamefont {A.}~\bibnamefont {Sen}},\ }\bibfield  {title} {\bibinfo {title} {{Dusty Plasma Experimental (DPEx) device for complex plasma experiments with flow}},\ }\href {https://doi.org/10.1063/1.4935608} {\bibfield  {journal} {\bibinfo  {journal} {Rev. Sci. Instrum.}\ }\textbf {\bibinfo {volume} {86}},\ \bibinfo {pages} {113503} (\bibinfo {year} {2015})}\BibitemShut {NoStop}%
\bibitem [{\citenamefont {Arora}\ \emph {et~al.}(2021)\citenamefont {Arora}, \citenamefont {Bandyopadhyay}, \citenamefont {Hariprasad},\ and\ \citenamefont {Sen}}]{Garima_PRE_2021}%
  \BibitemOpen
  \bibfield  {author} {\bibinfo {author} {\bibfnamefont {G.}~\bibnamefont {Arora}}, \bibinfo {author} {\bibfnamefont {P.}~\bibnamefont {Bandyopadhyay}}, \bibinfo {author} {\bibfnamefont {M.~G.}\ \bibnamefont {Hariprasad}},\ and\ \bibinfo {author} {\bibfnamefont {A.}~\bibnamefont {Sen}},\ }\bibfield  {title} {\bibinfo {title} {Experimental observation of pinned solitons in a flowing dusty plasma},\ }\href {https://doi.org/10.1103/PhysRevE.103.013201} {\bibfield  {journal} {\bibinfo  {journal} {Phys. Rev. E}\ }\textbf {\bibinfo {volume} {103}},\ \bibinfo {pages} {013201} (\bibinfo {year} {2021})}\BibitemShut {NoStop}%
\bibitem [{\citenamefont {Arora}\ \emph {et~al.}(2019)\citenamefont {Arora}, \citenamefont {Bandyopadhyay}, \citenamefont {Hariprasad},\ and\ \citenamefont {Sen}}]{Garima_POP_2019}%
  \BibitemOpen
  \bibfield  {author} {\bibinfo {author} {\bibfnamefont {G.}~\bibnamefont {Arora}}, \bibinfo {author} {\bibfnamefont {P.}~\bibnamefont {Bandyopadhyay}}, \bibinfo {author} {\bibfnamefont {M.~G.}\ \bibnamefont {Hariprasad}},\ and\ \bibinfo {author} {\bibfnamefont {A.}~\bibnamefont {Sen}},\ }\bibfield  {title} {\bibinfo {title} {Effect of size and shape of a moving charged object on the propagation characteristics of precursor solitons},\ }\href {https://doi.org/10.1063/1.5115313} {\bibfield  {journal} {\bibinfo  {journal} {Phys. Plasmas}\ }\textbf {\bibinfo {volume} {26}},\ \bibinfo {pages} {093701} (\bibinfo {year} {2019})}\BibitemShut {NoStop}%
\bibitem [{\citenamefont {Kumar}\ \emph {et~al.}(2024)\citenamefont {Kumar}, \citenamefont {Bandyopadhyay}, \citenamefont {Singh},\ and\ \citenamefont {Sen}}]{Krishan_POP_2024}%
  \BibitemOpen
  \bibfield  {author} {\bibinfo {author} {\bibfnamefont {K.}~\bibnamefont {Kumar}}, \bibinfo {author} {\bibfnamefont {P.}~\bibnamefont {Bandyopadhyay}}, \bibinfo {author} {\bibfnamefont {S.}~\bibnamefont {Singh}},\ and\ \bibinfo {author} {\bibfnamefont {A.}~\bibnamefont {Sen}},\ }\bibfield  {title} {\bibinfo {title} {{Excitation of cylindrical and spherical precursor solitons in a flowing dusty plasma: Experimental and simulation studies}},\ }\href {https://doi.org/10.1063/5.0177585} {\bibfield  {journal} {\bibinfo  {journal} {Phys. Plasmas}\ }\textbf {\bibinfo {volume} {31}},\ \bibinfo {pages} {023705} (\bibinfo {year} {2024})}\BibitemShut {NoStop}%
\bibitem [{\citenamefont {Tiwari}\ \emph {et~al.}(2012)\citenamefont {Tiwari}, \citenamefont {Das}, \citenamefont {Kaw},\ and\ \citenamefont {Sen}}]{Sanat_NJP_2012}%
  \BibitemOpen
  \bibfield  {author} {\bibinfo {author} {\bibfnamefont {S.~K.}\ \bibnamefont {Tiwari}}, \bibinfo {author} {\bibfnamefont {A.}~\bibnamefont {Das}}, \bibinfo {author} {\bibfnamefont {P.}~\bibnamefont {Kaw}},\ and\ \bibinfo {author} {\bibfnamefont {A.}~\bibnamefont {Sen}},\ }\bibfield  {title} {\bibinfo {title} {Observation of sharply peaked solitons in dusty plasma simulations},\ }\href {https://doi.org/10.1088/1367-2630/14/6/063008} {\bibfield  {journal} {\bibinfo  {journal} {New J. Phys.}\ }\textbf {\bibinfo {volume} {14}},\ \bibinfo {pages} {063008} (\bibinfo {year} {2012})}\BibitemShut {NoStop}%
\bibitem [{\citenamefont {Shukla}\ and\ \citenamefont {Mamun}(2001)}]{Shukla_IOP_2001}%
  \BibitemOpen
  \bibfield  {author} {\bibinfo {author} {\bibfnamefont {P.~K.}\ \bibnamefont {Shukla}}\ and\ \bibinfo {author} {\bibfnamefont {A.~A.}\ \bibnamefont {Mamun}},\ }\href@noop {} {\emph {\bibinfo {title} {Introduction to Dusty Plasma Physics}}}\ (\bibinfo  {publisher} {Institute of Physics, Bristol},\ \bibinfo {year} {2001})\BibitemShut {NoStop}%
\bibitem [{\citenamefont {Rao}\ \emph {et~al.}(1990)\citenamefont {Rao}, \citenamefont {Shukla},\ and\ \citenamefont {Yu}}]{Rao_PSS_1990}%
  \BibitemOpen
  \bibfield  {author} {\bibinfo {author} {\bibfnamefont {N.~N.}\ \bibnamefont {Rao}}, \bibinfo {author} {\bibfnamefont {P.~K.}\ \bibnamefont {Shukla}},\ and\ \bibinfo {author} {\bibfnamefont {M.~Y.}\ \bibnamefont {Yu}},\ }\bibfield  {title} {\bibinfo {title} {Dust-acoustic waves in dusty plasmas},\ }\href {https://doi.org/10.1016/0032-0633(90)90147-I} {\bibfield  {journal} {\bibinfo  {journal} {Planet. Space Sci.}\ }\textbf {\bibinfo {volume} {38}},\ \bibinfo {pages} {543} (\bibinfo {year} {1990})}\BibitemShut {NoStop}%
\bibitem [{\citenamefont {Kadomtsev}\ and\ \citenamefont {Petviashvili}(1970)}]{Kadomtsev_SPD_1970}%
  \BibitemOpen
  \bibfield  {author} {\bibinfo {author} {\bibfnamefont {B.~B.}\ \bibnamefont {Kadomtsev}}\ and\ \bibinfo {author} {\bibfnamefont {V.~I.}\ \bibnamefont {Petviashvili}},\ }\bibfield  {title} {\bibinfo {title} {On the stability of solitary waves in weakly dispersing media},\ }\href@noop {} {\bibfield  {journal} {\bibinfo  {journal} {Sov. Phys. Dokl.}\ }\textbf {\bibinfo {volume} {192}},\ \bibinfo {pages} {753} (\bibinfo {year} {1970})}\BibitemShut {NoStop}%
\bibitem [{\citenamefont {Cho}(2018)}]{Cho_PRE_2018}%
  \BibitemOpen
  \bibfield  {author} {\bibinfo {author} {\bibfnamefont {Y.}~\bibnamefont {Cho}},\ }\bibfield  {title} {\bibinfo {title} {{Stability of gravity-capillary solitary waves on shallow water based on the fifth-order Kadomtsev-Petviashvili equation}},\ }\href {https://doi.org/10.1103/PhysRevE.98.012213} {\bibfield  {journal} {\bibinfo  {journal} {Phys. Rev. E}\ }\textbf {\bibinfo {volume} {98}},\ \bibinfo {pages} {012213} (\bibinfo {year} {2018})}\BibitemShut {NoStop}%
\bibitem [{\citenamefont {Frycz}\ and\ \citenamefont {Infeld}(1989)}]{Frycz_PRL_1989}%
  \BibitemOpen
  \bibfield  {author} {\bibinfo {author} {\bibfnamefont {P.}~\bibnamefont {Frycz}}\ and\ \bibinfo {author} {\bibfnamefont {E.}~\bibnamefont {Infeld}},\ }\bibfield  {title} {\bibinfo {title} {Spontaneous transition from flat to cylindrical solitons},\ }\href {https://doi.org/10.1103/PhysRevLett.63.384} {\bibfield  {journal} {\bibinfo  {journal} {Phys. Rev. Lett.}\ }\textbf {\bibinfo {volume} {63}},\ \bibinfo {pages} {384} (\bibinfo {year} {1989})}\BibitemShut {NoStop}%
\bibitem [{\citenamefont {Infeld}\ \emph {et~al.}(1994)\citenamefont {Infeld}, \citenamefont {Senatorski},\ and\ \citenamefont {Skorupski}}]{Infeld_PRL_1994}%
  \BibitemOpen
  \bibfield  {author} {\bibinfo {author} {\bibfnamefont {E.}~\bibnamefont {Infeld}}, \bibinfo {author} {\bibfnamefont {A.}~\bibnamefont {Senatorski}},\ and\ \bibinfo {author} {\bibfnamefont {A.~A.}\ \bibnamefont {Skorupski}},\ }\bibfield  {title} {\bibinfo {title} {{Decay of Kadomtsev-Petviashvili solitons}},\ }\href {https://doi.org/10.1103/PhysRevLett.72.1345} {\bibfield  {journal} {\bibinfo  {journal} {Phys. Rev. Lett.}\ }\textbf {\bibinfo {volume} {72}},\ \bibinfo {pages} {1345} (\bibinfo {year} {1994})}\BibitemShut {NoStop}%
\bibitem [{\citenamefont {Senatorski}\ and\ \citenamefont {Infeld}(1996)}]{Senatorski_PRE_1996}%
  \BibitemOpen
  \bibfield  {author} {\bibinfo {author} {\bibfnamefont {A.}~\bibnamefont {Senatorski}}\ and\ \bibinfo {author} {\bibfnamefont {E.}~\bibnamefont {Infeld}},\ }\bibfield  {title} {\bibinfo {title} {{Simulations of Two-Dimensional Kadomtsev-Petviashvili Soliton Dynamics in Three-Dimensional Space}},\ }\href {https://doi.org/10.1103/PhysRevLett.77.2855} {\bibfield  {journal} {\bibinfo  {journal} {Phys. Rev. Lett.}\ }\textbf {\bibinfo {volume} {77}},\ \bibinfo {pages} {2855} (\bibinfo {year} {1996})}\BibitemShut {NoStop}%
\bibitem [{\citenamefont {Baronio}\ \emph {et~al.}(2016)\citenamefont {Baronio}, \citenamefont {Wabnitz},\ and\ \citenamefont {Kodama}}]{Baronio_PRL_2016}%
  \BibitemOpen
  \bibfield  {author} {\bibinfo {author} {\bibfnamefont {F.}~\bibnamefont {Baronio}}, \bibinfo {author} {\bibfnamefont {S.}~\bibnamefont {Wabnitz}},\ and\ \bibinfo {author} {\bibfnamefont {Y.}~\bibnamefont {Kodama}},\ }\bibfield  {title} {\bibinfo {title} {{Optical Kerr Spatiotemporal Dark-Lump Dynamics of Hydrodynamic Origin}},\ }\href {https://doi.org/10.1103/PhysRevLett.116.173901} {\bibfield  {journal} {\bibinfo  {journal} {Phys. Rev. Lett.}\ }\textbf {\bibinfo {volume} {116}},\ \bibinfo {pages} {173901} (\bibinfo {year} {2016})}\BibitemShut {NoStop}%
\bibitem [{\citenamefont {Cosme}\ and\ \citenamefont {Ter\ifmmode~\mbox{\c{c}}\else \c{c}\fi{}as}(2023)}]{Cosme_PRB_2023}%
  \BibitemOpen
  \bibfield  {author} {\bibinfo {author} {\bibfnamefont {P.}~\bibnamefont {Cosme}}\ and\ \bibinfo {author} {\bibfnamefont {H.}~\bibnamefont {Ter\ifmmode~\mbox{\c{c}}\else \c{c}\fi{}as}},\ }\bibfield  {title} {\bibinfo {title} {Nonlinear density waves on graphene electron fluids},\ }\href {https://doi.org/10.1103/PhysRevB.107.195432} {\bibfield  {journal} {\bibinfo  {journal} {Phys. Rev. B}\ }\textbf {\bibinfo {volume} {107}},\ \bibinfo {pages} {195432} (\bibinfo {year} {2023})}\BibitemShut {NoStop}%
\bibitem [{\citenamefont {Pakzad}\ and\ \citenamefont {Javidan}(2009)}]{Hamid_CSF_2009}%
  \BibitemOpen
  \bibfield  {author} {\bibinfo {author} {\bibfnamefont {H.~R.}\ \bibnamefont {Pakzad}}\ and\ \bibinfo {author} {\bibfnamefont {K.}~\bibnamefont {Javidan}},\ }\bibfield  {title} {\bibinfo {title} {Solitary waves in dusty plasmas with variable dust charge and two temperature ions},\ }\href {https://doi.org/https://doi.org/10.1016/j.chaos.2009.04.031} {\bibfield  {journal} {\bibinfo  {journal} {Chaos, Solitons \& Fractals}\ }\textbf {\bibinfo {volume} {42}},\ \bibinfo {pages} {2904} (\bibinfo {year} {2009})}\BibitemShut {NoStop}%
\bibitem [{\citenamefont {Dorranian}\ and\ \citenamefont {Sabetkar}(2012)}]{Davoud_POP_2012}%
  \BibitemOpen
  \bibfield  {author} {\bibinfo {author} {\bibfnamefont {D.}~\bibnamefont {Dorranian}}\ and\ \bibinfo {author} {\bibfnamefont {A.}~\bibnamefont {Sabetkar}},\ }\bibfield  {title} {\bibinfo {title} {{Dust acoustic solitary waves in a dusty plasma with two kinds of nonthermal ions at different temperatures}},\ }\href {https://doi.org/10.1063/1.3675883} {\bibfield  {journal} {\bibinfo  {journal} {Phys. Plasmas}\ }\textbf {\bibinfo {volume} {19}},\ \bibinfo {pages} {013702} (\bibinfo {year} {2012})}\BibitemShut {NoStop}%
\bibitem [{\citenamefont {Saini}\ \emph {et~al.}(2015)\citenamefont {Saini}, \citenamefont {Kaur},\ and\ \citenamefont {Gill}}]{Saini_ASR_2015}%
  \BibitemOpen
  \bibfield  {author} {\bibinfo {author} {\bibfnamefont {N.}~\bibnamefont {Saini}}, \bibinfo {author} {\bibfnamefont {N.}~\bibnamefont {Kaur}},\ and\ \bibinfo {author} {\bibfnamefont {T.}~\bibnamefont {Gill}},\ }\bibfield  {title} {\bibinfo {title} {{Dust acoustic solitary waves of Kadomstev–Petviashvili (KP) equation in superthermal dusty plasma}},\ }\href {https://doi.org/https://doi.org/10.1016/j.asr.2015.03.013} {\bibfield  {journal} {\bibinfo  {journal} {Adv. Space Res.}\ }\textbf {\bibinfo {volume} {55}},\ \bibinfo {pages} {2873} (\bibinfo {year} {2015})}\BibitemShut {NoStop}%
\bibitem [{\citenamefont {Mir}\ \emph {et~al.}(2022)\citenamefont {Mir}, \citenamefont {Tiwari},\ and\ \citenamefont {Sen}}]{Ajaz_POP_2022}%
  \BibitemOpen
  \bibfield  {author} {\bibinfo {author} {\bibfnamefont {A.}~\bibnamefont {Mir}}, \bibinfo {author} {\bibfnamefont {S.}~\bibnamefont {Tiwari}},\ and\ \bibinfo {author} {\bibfnamefont {A.}~\bibnamefont {Sen}},\ }\bibfield  {title} {\bibinfo {title} {{Bispectral analysis of nonlinear mixing in a periodically driven Korteweg–de Vries system}},\ }\href {https://doi.org/10.1063/5.0077638} {\bibfield  {journal} {\bibinfo  {journal} {Phys. Plasmas}\ }\textbf {\bibinfo {volume} {29}},\ \bibinfo {pages} {032303} (\bibinfo {year} {2022})}\BibitemShut {NoStop}%
\bibitem [{\citenamefont {Klein}\ \emph {et~al.}(2007)\citenamefont {Klein}, \citenamefont {Sparber},\ and\ \citenamefont {Markowich}}]{Klein_JNS_2007}%
  \BibitemOpen
  \bibfield  {author} {\bibinfo {author} {\bibfnamefont {C.}~\bibnamefont {Klein}}, \bibinfo {author} {\bibfnamefont {C.}~\bibnamefont {Sparber}},\ and\ \bibinfo {author} {\bibfnamefont {P.}~\bibnamefont {Markowich}},\ }\bibfield  {title} {\bibinfo {title} {{Numerical study of oscillatory regimes in the Kadomtsev--Petviashvili equation}},\ }\href@noop {} {\bibfield  {journal} {\bibinfo  {journal} {J. Nonlinear Sci.}\ }\textbf {\bibinfo {volume} {17}},\ \bibinfo {pages} {429} (\bibinfo {year} {2007})}\BibitemShut {NoStop}%
\bibitem [{\citenamefont {Kumar}\ \emph {et~al.}(2023)\citenamefont {Kumar}, \citenamefont {Bandyopadhyay}, \citenamefont {Singh},\ and\ \citenamefont {Sen}}]{Krishan_POP_2023}%
  \BibitemOpen
  \bibfield  {author} {\bibinfo {author} {\bibfnamefont {K.}~\bibnamefont {Kumar}}, \bibinfo {author} {\bibfnamefont {P.}~\bibnamefont {Bandyopadhyay}}, \bibinfo {author} {\bibfnamefont {S.}~\bibnamefont {Singh}},\ and\ \bibinfo {author} {\bibfnamefont {A.}~\bibnamefont {Sen}},\ }\bibfield  {title} {\bibinfo {title} {{Interaction of a precursor soliton with wake structure in a flowing dusty plasma}},\ }\href {https://doi.org/10.1063/5.0149355} {\bibfield  {journal} {\bibinfo  {journal} {Phys. Plasmas}\ }\textbf {\bibinfo {volume} {30}},\ \bibinfo {pages} {073701} (\bibinfo {year} {2023})}\BibitemShut {NoStop}%
\bibitem [{\citenamefont {Mir}\ \emph {et~al.}(2023)\citenamefont {Mir}, \citenamefont {Tiwari}, \citenamefont {Sen}, \citenamefont {Crabtree}, \citenamefont {Ganguli},\ and\ \citenamefont {Goree}}]{Ajaz_PRE_2023}%
  \BibitemOpen
  \bibfield  {author} {\bibinfo {author} {\bibfnamefont {A.}~\bibnamefont {Mir}}, \bibinfo {author} {\bibfnamefont {S.}~\bibnamefont {Tiwari}}, \bibinfo {author} {\bibfnamefont {A.}~\bibnamefont {Sen}}, \bibinfo {author} {\bibfnamefont {C.}~\bibnamefont {Crabtree}}, \bibinfo {author} {\bibfnamefont {G.}~\bibnamefont {Ganguli}},\ and\ \bibinfo {author} {\bibfnamefont {J.}~\bibnamefont {Goree}},\ }\bibfield  {title} {\bibinfo {title} {{Synchronization of dust acoustic waves in a forced Korteweg--de Vries--Burgers model}},\ }\href {https://doi.org/10.1103/PhysRevE.107.035202} {\bibfield  {journal} {\bibinfo  {journal} {Phys. Rev. E}\ }\textbf {\bibinfo {volume} {107}},\ \bibinfo {pages} {035202} (\bibinfo {year} {2023})}\BibitemShut {NoStop}%
\bibitem [{\citenamefont {Ott}\ and\ \citenamefont {Sudan}(1969)}]{Ott_POF_1969}%
  \BibitemOpen
  \bibfield  {author} {\bibinfo {author} {\bibfnamefont {E.}~\bibnamefont {Ott}}\ and\ \bibinfo {author} {\bibfnamefont {R.~N.}\ \bibnamefont {Sudan}},\ }\bibfield  {title} {\bibinfo {title} {{Nonlinear Theory of Ion Acoustic Waves with Landau Damping}},\ }\href {https://doi.org/10.1063/1.1692358} {\bibfield  {journal} {\bibinfo  {journal} {Phys. Fluids}\ }\textbf {\bibinfo {volume} {12}},\ \bibinfo {pages} {2388} (\bibinfo {year} {1969})}\BibitemShut {NoStop}%
\end{thebibliography}%
%%%%%%%%%%%%%%%%%%%%%%%%%%%%%%%%%%%%%%%%%%%%%%%%%%%%%%%%%%%%%%%%%%%%%%%%%%%%%%%%
%%%%%%%%%%%%%%%%%%%%%%%%%%%%%%%%%%%%%%%%%%%%%%%%%%%%%%%%%%%%%%%%%%%%%%%%%%%%%%%%
\end{document}